\documentclass[final,12pt,authoryear]{elsarticle}
\journal{Journal of the Mechanics and Physics of Solids}

\usepackage{amsfonts,amssymb,amsmath,amsthm,autobreak,cancel,
    dsfont,mathtools,mathrsfs,siunitx,upgreek}

\usepackage{booktabs,longtable,dcolumn,makecell,
    multicol,multirow,tabularx,xltabular,rotating}

\usepackage[pagewise,mathlines]{lineno}     % line numbers for drafting
\usepackage[ruled]{algorithm2e}             % to manage algorithm environment
\usepackage{caption,subcaption}             % to manage captions
\usepackage{color}                          % color related packages
\usepackage{enumitem}                       % to manage list environment
\usepackage{etoolbox}
\usepackage{float}                          % to manage floating environment
\usepackage{graphicx,wrapfig}               % to manage images
\usepackage{geometry}
\usepackage[dvipsnames]{xcolor}             % to manage colors
\usepackage{hyperref}                       % for hyperlinks
\usepackage[all]{hypcap}                    % for captions on the side of figures
\usepackage{ifthen}                         % if-then statement in algorithm
\usepackage{lscape}                         % landscape mode
\usepackage{listings,minted}                % to include codes
\usepackage{csquotes}                       % for managing quotes
\usepackage{marvosym}
\usepackage{parskip}                        % paragraph spacing package
\usepackage{seqsplit}                       % splits long character sequence
\usepackage[rightcaption]{sidecap}          % for sideway captions
\usepackage{soul}
\usepackage[absolute]{textpos}              % to position text
\usepackage{tikz}                           % package for drawing

\hypersetup{linktocpage, unicode, linktoc=all, colorlinks=true, 
    citecolor=blue, filecolor=blue, linkcolor=blue, urlcolor=blue}
\theoremstyle{definition}
\newtheorem{remark}{Remark}
\counterwithin{remark}{section}
\AtBeginEnvironment{remark}{\vspace{\baselineskip}}
\newcommand{\dC}{$^{\circ}$C}
\DeclareMathOperator{\T}{{\top}}

\newcommand{\vect}[1]{\mathbf{#1}}

\newcommand{\barVect}[1]{\overline{\mathbf{#1}}}

\DeclareMathOperator{\tr}{tr}

\DeclareMathOperator{\dev}{dev}

\DeclareMathOperator{\eq}{eq}
\DeclareMathOperator{\neql}{neq}
\DeclareMathOperator{\vol}{vol}

\DeclareMathOperator{\vsigma}{\boldsymbol{\upsigma}}
\DeclareMathOperator{\vtau}{\boldsymbol{\uptau}}

\makeatletter
\DeclareRobustCommand\bigop[2][1]{%
  \mathop{\vphantom{\sum}\mathpalette\bigop@{{#1}{#2}}}\slimits@
}
\newcommand{\bigop@}[2]{\bigop@@#1#2}
\newcommand{\bigop@@}[3]{%
  \vcenter{%
    \sbox\z@{$#1\sum$}%
    \hbox{\resizebox{\ifx#1\displaystyle#2\fi\dimexpr\ht\z@+\dp\z@}{!}{$\m@th#3$}}%
  }%
}
\makeatother

\makeatletter
\renewcommand*\env@matrix[1][*\c@MaxMatrixCols c]{%
  \hskip -\arraycolsep
  \let\@ifnextchar\new@ifnextchar
  \array{#1}}
\makeatother

\begin{document}
% \linenumbers

%%%%%% FRONT MATTER
\begin{frontmatter}

%%%%% ARTICLE TITLE
\title{A finite viscoelastic constitutive model for low to high strain rate response of elastomers with application of strain rate-induced glass transition}

%%%%% AUTHORS
\author[1]{Bibekananda Datta}
\author[1]{Sushan Nakarmi}
\author[1]{Nitin P. Daphalapurkar\corref{cor}}     \ead{nitin@lanl.gov}
\cortext[cor]{Corresponding author}

%%%%%% AFFILIATIONS
\affiliation[1]{organization={Theoretical Division, Los Alamos National Laboratory},
            city={Los Alamos},
            postcode={87545}, 
            state={NM},
            country={USA}}

%%%% ABSTRACT
\begin{abstract}

Amorphous elastomers exhibit significant rate-stiffening and unique viscous flow characteristics across a wide range of strain rates, often undergoing glass transition above a strain rate threshold. We have developed a thermodynamically-consistent and micromechanically-inspired constitutive model for soft elastomeric materials to capture the rate-dependent stress-strain behavior and hysteresis when subjected to low to high strain rates. Our proposed constitutive model encapsulates the viscous flow of materials through molecular motion at low strain rates and local rearrangement and alignment of the molecules trying to overcome the intermolecular resistance at high strain rates, essentially covering the glass transition. We applied our constitutive model to uniaxial compression experiments performed at low and high strain rates for polyborosiloxane (PBS) to identify the material parameters, and subsequently, performed numerical simulations of single and multi-cycle compression, stress relaxation, and small amplitude oscillatory tension-compression. Our analyses indicate that the model predicts higher total energy dissipation with increasing strain rate; however, dissipation associated with molecular relaxation decreases (forming a cusp) because, beyond a crossover strain rate, intermolecular rearrangement and alignment become dominant, which is consistent with the onset of the glass transition. For cyclic loading-unloading, we observed that dissipation over a cycle remains constant at low strain rates but decreases non-monotonically at high strain rates before becoming constant, with the peak stress over the cycle becoming higher, which can be interpreted as more loading being carried elastically by the polymer network as the intermolecular rearrangement process occurs. Additionally, our model was able to predict the qualitative nature of the storage modulus and loss modulus in the limit of small strain over a wide range of frequency sweeps.

\end{abstract}

%%%%% KEYWORDS

\end{frontmatter}

%%%%%% MAIN MATTER
\section{Introduction}

Amorphous elastomers have been widely used for decades in engineering applications such as vibration isolation, shock absorption, and impact protection because of their large stretchability, high compliance, and viscous damping characteristics. From an application perspective, it is expected that energy-absorbing elastomeric structures will be subjected to mechanical loadings over a wide range of strain rate regimes: (a) quasi-static $(10^{-6}-10^{-1}$ s$^{-1}$), (b) moderate ($10^{-1}-10^{1}$ s$^{-1}$), (c) high ($10^2-10^4$ s$^{-1}$), and (d) shock loading ($10^5-10^7$ s$^{-1}$). Different micro-mechanisms with distinct characteristic relaxation times cause markedly varied macroscopic stress-strain responses in different strain rate regimes. For example, in the quasi-static regime, the molecular-level kinetic events or the dynamics of polymer chains give rise to the viscoelastic nature of elastomers with strong rate-dependency and hysteresis. However, as the strain rate is increased beyond a threshold, rubbery elastomers undergo the glass transition. This makes the polymer chains \enquote{frozen} in their configurations with little to no mobility, resulting in nonlinear viscous flow and subsequent hardening followed by yielding. These unique characteristics of amorphous elastomers across a wide range of strain rates nullify the possibility of using a stand-alone hyperelastic model to study rate-dependency and hysteresis behaviors and demand sophisticated viscoelastic models. In this work, we develop a thermodynamically-consistent, mechanism‑based constitutive model that provides accurate prediction of the mechanical response of amorphous elastomers exhibiting rate stiffening and glass transition across six decades of strain rates, spanning quasi‑static to high strain rate dynamic regimes ($10^{-3}-10^3$ s$^{-1}$).

The continuum theories of viscoelasticity can be classified into two distinct approaches. The first approach includes the early works on the time‑dependent behavior of materials, which can be traced back to the rheologically inspired, convolutional integral‑based phenomenological models developed by Kelvin, Maxwell, and Voigt in the small strain regime utilizing \emph{stress-type} internal variables \citep{christensenTheoryViscoelasticityIntroduction1982}. Traditional linear viscoelastic models, albeit intended for small-strain quasi-static applications, have been applied in modeling large deformation of polymers at high strain rates. However, they often require additional modification, such as incorporation of damage mechanics \citep{chenApplicationLinearViscoelastic2020}, or calibration of a large set of material parameters \citep{trivediSimpleRateTemperature2020}. As the engineering applications of polymers and the analyses of soft biological tissues became common in the 20th century, linear viscoelastic models were extended for large deformation with the initial focus of applications in the quasi-static regime \citep{simoFullyThreedimensionalFinitestrain1987,govindjeeMullinsEffectStrain1992,fungBiomechanicsMechanicalProperties1993,letallecThreedimensionalIncompressibleViscoelasticity1993,holzapfelNewViscoelasticConstitutive1996,kaliskeFormulationImplementationThreedimensional1997}. These so-called quasi-linear or visco-hyperelastic models resolved some of the issues associated with the application of linear viscoelastic models for high strain rate cases and were subsequently adapted for different types of rubbers \citep{yangViscohyperelasticApproachModelling2000,shimViscohyperelasticConstitutiveModel2004,hoofattIntegralbasedConstitutiveEquation2007,pouriayevaliConstitutiveDescriptionElastomer2012}. However, because of their inherent indifference to underlying mechanisms, so far, they have not been applied in modeling temperature or strain rate-induced glass transition.

The second approach, motivated by the multiplicative plasticity theory \citep{leeElasticplasticDeformationFinite1969}, was developed to describe the finite viscoelastic behaviors of materials using viscous strain as the internal variable in a thermodynamically-consistent manner \citep{sidoroffModeleViscoelastiqueNon1974,lublinerModelRubberViscoelasticity1985,lionPhysicallyBasedMethod1997,reeseTheoryFiniteViscoelasticity1998,bergstromConstitutiveModelingLarge1998}. In contrast to the convolutional integral approach, this approach allows incorporating micromechanically-motivated evolution laws for viscous strain, potentially covering a wide range of relaxation times. For instance, \cite{reeseMicromechanicallyMotivatedMaterial2003} adopted the transient network theory \citep{tanakaViscoelasticPropertiesPhysically1992,tanakaViscoelasticPropertiesPhysically1992a}, derived based on the kinetics of formation and breakage of polymer chain segments \citep{greenNewApproachTheory1946}, within the finite viscoelastic framework developed earlier by the same author \citep{reeseTheoryFiniteViscoelasticity1998}. In a similar manner, \cite{bergstromConstitutiveModelingLarge1998} and \cite{mieheMicromacroApproachRubberlike2005} adopted the reptation motion-based polymer chain dynamics model \citep{degennesReptationPolymerChain1971,doiTheoryPolymerDynamics1986} in their own continuum mechanics frameworks to describe the finite nonlinear viscoelasticity of rubber-like materials. Except for the work of \cite{quintavallaExtensionBergstromboyceModel2004}, who successfully applied the Bergstr\"{o}m-Boyce model \citep{bergstromConstitutiveModelingLarge1998} for high strain rate deformation of polybutadiene, literature concerning mechanistic modeling of large deformation of amorphous elastomers at high strain rates is largely absent.

\cite{sarvaStressStrainBehavior2007,yiLargeDeformationRatedependent2006,hoofattIntegralbasedConstitutiveEquation2007,konaleLargeDeformationModel2023} reported strain rate-induced glass transition of different elastomeric materials. Among these authors, \cite{konaleLargeDeformationModel2023} corroborated their experimental findings for polyborosiloxane (PBS) with a semi-physical viscoplastic constitutive model. However, their proposed constitutive model fails to capture the yielding and hardening behaviors observed at high strain rates, which are the characteristic traits of the glassy state of amorphous polymers. Unlike the rubbery elastomers, glassy polymers exhibit stiff elastic behavior in the early stage of deformation due to the lack of time for the chains to be mobile, and the viscous flow and hardening past yielding are attributed to the local rearrangement of the molecules. Thus far, researchers utilized both phenomenological \citep{anandModelingMicroindentationResponse2006} and mechanistic models \citep{boyceLargeInelasticDeformation1988,boyceConstitutiveModelFinite2000} to describe the viscous flow in glassy materials. Pioneering work by \cite{mullikenMechanicsRatedependentElastic2006} leveraged the latter approach to develop a model for strain rate-induced $\alpha$- and $\beta$-transitions of glassy polycarbonate (PC) and poly(methyl methacrylate) (PMMA). However, despite the existence of an extensive literature on the temperature-driven glass transition of amorphous polymers \citep{dupaixConstitutiveModelingFinite2007,nguyenThermoviscoelasticModelAmorphous2008,anandThermomechanicallyCoupledTheory2009,amesThermomechanicallyCoupledTheory2009,xiaoModelingGlassTransition2013}, there has not been any attempt to develop a mechanistic constitutive model for elastomers over a wide range of strain rates that can capture strain rate-induced glass transition.

In this work, we propose a mechanistically-motivated constitutive model for finite nonlinear viscoelastic deformation of elastomers across a wide range of strain rates to address strain rate-induced glass transition. Micro-mechanisms of deformation were adopted within the finite deformation viscoelastic theory proposed by \cite{reeseTheoryFiniteViscoelasticity1998}, briefly summarized in Section \ref{sec:finite-viscoelasticity-theory}. At low strain rates, the viscous flow of the rubbery elastomer is described using the reptation motion-based dynamics of polymer chains \citep{bergstromConstitutiveModelingLarge1998}, and at high strain rates, the viscous flow is described by the Ree-Eyring flow rule with an evolving yield surface \citep{nguyenThermoviscoelasticModelAmorphous2008}. The details of the constitutive model and procedure to identify the material parameters are given in Section \ref{sec:constitutive-model}. In Section \ref{sec:results-discussion}, we quantitatively demonstrate the characteristic features of our proposed constitutive model using multiple numerical examples. We studied the strain rate sensitivity, energy dissipation behavior, stress relaxation, and small-strain dynamic moduli at different strain rates. Associated limitations of our proposed model, possible improvements, and future directions are presented in Section \ref{sec:conclusions}.

\section{A brief review of finite nonlinear viscoelastic theory}
\label{sec:finite-viscoelasticity-theory}

Motivated by the associated plasticity theory \citep{simoAssociativeCoupledThermoplasticity1992,simoAlgorithmsStaticDynamic1992}, \cite{reeseTheoryFiniteViscoelasticity1998} presented a thermodynamically-consistent finite viscoelastic theory without resorting to any underlying mechanisms that can be attributed to specific materials. However, their theory has later been adopted to describe mechanism-specific finite viscoelasticity of rubbers \citep{reeseMicromechanicallyMotivatedMaterial2003,dalBergstromBoyceModel2009} as well as glassy polymers \citep{nguyenThermoviscoelasticModelAmorphous2008,xiaoModelingGlassTransition2013}. In this section, we briefly outline the finite deformation kinematics and thermodynamic restrictions as a premise to set up our constitutive model in Section \ref{sec:constitutive-model}.

\begin{figure}[!ht]
    \centering
    \includegraphics[width=\textwidth, page=1, clip, trim=2.25cm 7cm 2cm 7.25cm]{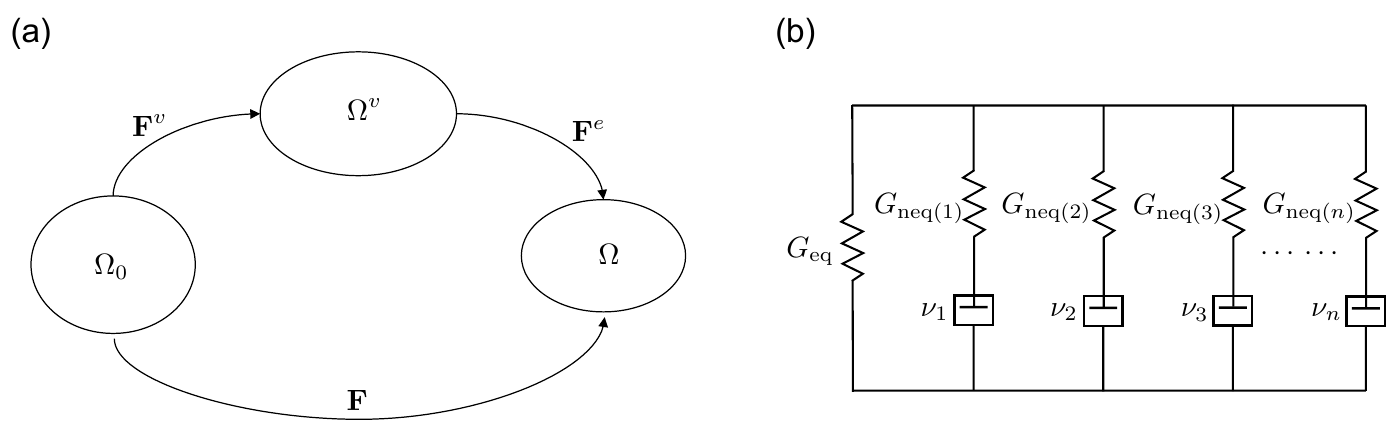}
    \caption{(a) Multiplicative decomposition of the deformation gradient into elastic and viscous parts, (b) a simple rheological representation of the generalized Maxwell model.}
    \label{fig:kinematics-and-rheology}
\end{figure}

\subsection{Multiplicative finite deformation kinematics}

As shown in Figure \ref{fig:kinematics-and-rheology}(a), the primary feature of the Reese-Govindjee viscoelasticity theory \citep{reeseTheoryFiniteViscoelasticity1998} is the multiplicative split of the total deformation gradient, $\vect{F}$, into elastic parts, $\vect{F}^e_k$, and inelastic viscous parts, $\vect{F}^v_k$, as follows \citep{leeElasticplasticDeformationFinite1969,sidoroffModeleViscoelastiqueNon1974}
\begin{equation}    \label{eq:multiplicative-decomposition}
    \vect{F} = \vect{F}^e_k \vect{F}^v_k,
\end{equation}
where the subscript, $(\bullet)_k$, represents a Maxwell branch in a standard linear solid-like rheological configuration, as illustrated in Figure \ref{fig:kinematics-and-rheology}(b). In our adaptation, we assume that the spring and dashpot in each Maxwell element represent a unique kind of resistance and viscous flow mechanism, respectively. In this work, we are specifically interested in prescribing viscous mechanisms that can capture the rubbery and glassy behaviors in parallel rheological configuration at low and high strain rates, respectively. Consequently, they can also describe the strain rate-induced glass transition of the amorphous elastomers.

As is standard in continuum mechanics, we can define the total left Cauchy-Green tensor, $\vect{b}$, and its elastic counterparts, $\vect{b}^e_k$, as follows:
\begin{equation}
\begin{gathered}
    \vect{b} = \vect{F} \vect{F}^{\T} 
    = \vect{F}^e_k {\vect{b}^v_k}^{\T} {\vect{F}^e_k}^{\T}
    \qquad
    \text{and} \quad
    \vect{b}^e_k = \vect{F}^e_k {\vect{F}^e_k}^{\T}
    = \vect{F} {\vect{C}^v_k}^{-1} \vect{F}^{\T},
\end{gathered}
\end{equation}
where $\vect{b}^v_k$ and $\vect{C}^v_k$ are the left and right viscous Cauchy-Green deformation tensors, respectively, associated with the $k-$th viscous mechanism. By utilizing the multiplicative split of the deformation gradient introduced earlier in Eq. \eqref{eq:multiplicative-decomposition}, one can decompose the velocity gradient, $\vect{l}$, additively as follows:
\begin{equation}    \label{eq:split-of-velocity-grad}
\begin{aligned}
    \vect{l} = \dot{\vect{F}} \vect{F}^{-1} = \vect{l}^e_k + \hat{\vect{l}}^v_k,
\end{aligned}
\end{equation}
where
\begin{equation}
    \vect{l}^e_k = \dot{\vect{F}}^e_k {\vect{F}^e_k}^{-1},
    \qquad
    \vect{l}^v_k = \dot{\vect{F}}^v_k {\vect{F}^v_k}^{-1},
    \qquad \text{and} \qquad 
    \hat{\vect{l}}^v_k = \vect{F}^e_k \vect{l}^v_k {\vect{F}^e_k}^{-1}
\end{equation}
represent the elastic, viscous, and spatial viscous velocity gradients, respectively. As is standard, these tensors can be further decomposed into their own stretch and spin tensor counterparts in the following manner:
\begin{equation}
    \vect{l}^e_k  = \vect{d}^e_k + \vect{w}^e_k,
    \qquad
    \vect{l}^v_k  = \vect{d}^v_k + \vect{w}^v_k,
    \qquad \text{and} \qquad
    \hat{\vect{l}}^v_k = \hat{\vect{d}}^v_k + \hat{\vect{w}}^v_k,
\end{equation}
where $\vect{d}^{(\bullet)}_k$ is the symmetric part and $\vect{w}^{(\bullet)}_k$ is the anti-symmetric part, respectively, of the corresponding velocity gradient, $\vect{l}^{(\bullet)}_k$.

\subsection{Constitutive restrictions}

For an isothermal process, where $\vect{C}^e_k$ is the internal variable associated with the $k-$th viscous mechanism, the second law of thermodynamics dictates the dissipation inequality to appear as \citep{colemanThermodynamicsInternalState1967}:
\begin{equation}    \label{eq:dissipation-inequality}
    \left( \vect{S} - 2 \frac{\partial \Psi}{\partial \vect{C}} \right):\frac{1}{2} \dot{\vect{C}}
    - \sum_{k=1}^n \frac{\partial \Psi}{\partial \vect{C}^e_k} : \dot{\vect{C}}^e_k
    \geq 0.
\end{equation}
where $\vect{S}$ is the second Piola-Kirchhoff stress, and $\Psi$ is the total Helmholtz free energy density which can be considered to be an additive decomposition of an equilibrium part, $\Psi_{\eq}$, and multiple non-equilibrium parts, $\Psi_{\neql(k)}$, as follows:
\begin{equation} \label{eq:free-energy-split}
    \Psi (\vect{C}; \vect{C}^e_k)
    = \Psi_{\eq}(\vect{C}) + \sum_{k=1}^n \Psi_{\neql(k)}(\vect{C}^e_k).
\end{equation}
With the aid of the kinematic description of finite viscoelasticity in Eq. \eqref{eq:multiplicative-decomposition} and the free energy density ansatz in Eq. \eqref{eq:free-energy-split}, the dissipation inequality Eq. \eqref{eq:dissipation-inequality} can be written as:
\begin{equation}    \label{eq:dissipation-inequality-revised}
    \left( \vect{S} - 2 \frac{\partial \Psi_{\eq}}{\partial \vect{C}} 
    - \sum_{k=1}^n 2 {\vect{F}^v_k}^{-1}  \frac{\partial \Psi_{\neql(k)}}{\partial \vect{C}^e_k}  {\vect{F}^v_k}^{-\T} \right)
    : \frac{1}{2} \dot{\vect{C}}
    - \sum_{k=1}^n \vtau_{\neql(k)} : \frac{1}{2} \mathcal{L}_v \vect{b}^e_k {\vect{b}^e_k}^{-1} \geq 0,
\end{equation}
where $\mathcal{L}_v$ is the Lie-time derivative. Hereafter, the application of Coleman-Noll argument \citep{colemanThermodynamicsElasticMaterials1963} and subsequently performing a push-forward operation on the second Piola-Kirchhoff stress, $\vect{S}$, yields the thermodynamic restrictions on the Cauchy stress, $\vsigma$, as follows:
\begin{equation}    \label{eq:thermodynamic-restrictions}
    \vsigma 
    =  \underbrace{ J^{-1} \vect{F} \left( 2 \frac{\partial \Psi_{\eq}} {\partial \vect{C}} \right) \vect{F}^{\T}}_{\vsigma_{\eq}}
    + \sum_{k=1}^n \underbrace{ J^{-1} \vect{F}^e_k \left( 2 \frac{\partial \Psi_{\neql(k)}} {\partial \vect{C}^e_k} \right) {\vect{F}^e_k}^{\T}}_{\vsigma_{\neql(k)}},
\end{equation}
where $\vsigma_{\eq}$ is the equilibrium Cauchy stress associated with the static polymer network response, and $\vsigma_{\neql(k)}$ is known as the non-equilibrium stress or overstress, and is associated with each viscous mechanism, $k$. To satisfy the thermodynamic consistency of the residual dissipation part in Eq. \eqref{eq:dissipation-inequality-revised}, \cite{reeseTheoryFiniteViscoelasticity1998} prescribed the following constitutive law:
\begin{equation}    \label{eq:RG-viscous-constitutive-form}
    -\frac{1}{2} \mathcal{L}_v \vect{b}^e_k {\vect{b}^e_k}^{-1} 
    = \hat{\vect{d}}^v_k
    = \left(\boldsymbol{\upnu}_k^{-1}: \vtau_{\neql(k)} \right),
\end{equation}
where $\vtau_{\neql(k)}= J \vsigma_{\neql(k)}$ is the non-equilibrium Kirchhoff stress and $\boldsymbol{\upnu}_k$ is a fourth-order isotropic positive definite tensor which is constitutively prescribed for the $k-$th viscous mechanism. \cite{reeseTheoryFiniteViscoelasticity1998} adopted the following form for the viscosity tensor:
\begin{equation}    \label{eq:RG-inverse-viscosity-tensor}
    \boldsymbol{\upnu}^{-1}_k
    = \frac{1}{2 \nu^{\dev}_k} \left( \mathbb{I} - \frac{1}{3} \mathds{1} \otimes \mathds{1} \right) 
    + \frac{1}{9 \nu^{\vol}_k} \mathds{1} \otimes \mathds{1},
\end{equation}
where $\nu^{\dev}_k$ and $\nu^{\vol}_k$ are the nonlinear deviatoric and volumetric viscosity functions, respectively, for the $k-$th viscous mechanism.

\subsection{Volume-decoupled extension of the theory}

Since the rubbery elastomers are known to be nearly-incompressible whereas the glassy polymers are considered to be fairly compressible, it is advantageous to leverage the definition of isochoric deformation gradient, $\barVect{F} = J^{-1/3} \vect{F}$, to further split the free energy density, $\Psi$, in the following form:
\begin{equation}    \label{eq:decoupled-free-energy-split}
    \Psi(\barVect{C}, J; \vect{C}^e_k, J^e_k) 
    = \Psi_{\eq}^{\dev} (\barVect{C}) + \Psi_{\eq}^{\vol}(J)
    + \sum_{k=1}^n \Psi_{\neql(k)}^{\dev}(\barVect{C}^e_k) 
    + \sum_{k=1}^n \Psi_{\neql(k)}^{\vol}(J^e_k),
\end{equation}
where $\Psi_{\eq}^{\dev}$ and $\Psi_{\eq}^{\vol}$ represent the deviatoric and volumetric parts of the equilibrium free energy density with $\barVect{C} = \barVect{F}^{\T} \barVect{F}$ representing the right isochoric Cauchy-Green tensor. The deviatoric and volumetric parts of the free energy density pertaining to each viscous mechanism, $k$, are represented by $\Psi_{\neql(k)}^{\dev}$ and $\Psi_{\neql(k)}^{\vol}$, respectively, with right isochoric elastic Cauchy-Green tensor, $\barVect{C}^e_k = {\barVect{F}^e_k}^{\T} \barVect{F}^e_k$, and elastic volume change, $J^e_k = \det(\vect{F}^e_k)$. As a consequence of the free energy density ansatz introduced in Eq. \eqref{eq:decoupled-free-energy-split}, Cauchy stress, $\vsigma$, can now be written as follows:
\begin{equation}    \label{eq:additive-split-stress}
    \vsigma = \vsigma_{\eq}^{\dev} + \vsigma_{\eq}^{\vol} 
    + \sum_{k=1}^n \vsigma_{\neql(k)}^{\dev} + \sum_{k=1}^n \vsigma_{\neql(k)}^{\vol},
\end{equation}
with the following thermodynamic restrictions:
\begin{equation} \label{eq:thermodynamic-restriction-stress-split}
\begin{aligned}      
    & \vsigma_{\eq}^{\dev}
    && =  J^{-1} \barVect{F} \left( \mathbb{P} : 2 \frac{\partial \Psi_{\eq}^{\dev}} {\partial \barVect{C}} \right) \barVect{F}^{\T}, 
    \quad
    && \vsigma_{\eq}^{\vol}
    && = \frac{\partial \Psi_{\eq}^{\vol}} {\partial J} \mathds{1}, \\
    & \vsigma_{\neql(k)}^{\dev}
    && = J^{-1} \left[ \barVect{F}^e_k \left( \mathbb{P}_{e(k)} : 2 \frac{\partial \Psi_{\neql(k)}^{\dev}} {\partial \barVect{C}^e_k}  \right) {\barVect{F}^e_k}^{\T} \right],
    \quad
    && \vsigma_{\neql(k)}^{\vol}
    && = \frac{J^e_k}{J} \frac{\partial \Psi_{\neql(k)}^{\vol}} {\partial J^e_k}  \mathds{1},
\end{aligned}
\end{equation}
with
\begin{equation}
\begin{aligned}
    & \mathbb{P}        && = \mathbb{I} - \frac{1}{3} \vect{C}^{-1} \otimes \vect{C}
    && \quad \text{and} \quad
    && \mathbb{P}_{e(k)} && = \mathbb{I} - \frac{1}{3} {\vect{C}^e_k}^{-1} \otimes \vect{C}^e_k
\end{aligned}
\end{equation}
defined as the total and elastic projection tensors related to the $k-$th viscous mechanism, respectively. The kinematic relation, $-\frac{1}{2} \mathcal{L}_v \vect{b}^e_k {\vect{b}^e_k}^{-1} = \hat{\vect{d}}^v_k$, from Eq. \eqref{eq:RG-viscous-constitutive-form}, and the deviatoric and volumetric split of Cauchy stress, from Eq. \eqref{eq:additive-split-stress}, allow us to write the residual dissipation Eq. \eqref{eq:dissipation-inequality-revised} in the following form:
\begin{equation}    \label{eq:residual-dissipation-split} 
    \sum_{k=1}^n \left( \vtau_{\neql(k)}^{\dev} : \hat{\vect{d}}^{v,\dev}_k
    + \vtau_{\neql(k)}^{\vol} : \hat{\vect{d}}^{v,\vol}_k \right) \geq 0.
\end{equation}
where $\vtau_{\neql(k)}^{\dev} = J \vsigma_{\neql(k)}^{\dev}$ and $\vtau_{\neql(k)}^{\vol} = J \vsigma_{\neql(k)}^{\vol}$ are the deviatoric and volumetric parts of the non-equilibrium Kirchhoff stress, respectively, and $\hat{\vect{d}}^{v,\dev}_k$ and $\hat{\vect{d}}^{v,\vol}_k$ represent the deviatoric and volumetric parts of the spatial viscous stretch tensors, respectively, such that the kinematic relation, $\hat{\vect{d}}^v_k = \hat{\vect{d}}^{v,\dev}_k + \hat{\vect{d}}^{v,\vol}_k$, holds true for each $k-$th viscous mechanism. We can now prescribe the following forms of viscous evolution laws for the deviatoric and volumetric parts:
\begin{equation}    \label{eq:stretch-rate-constitutive-law}
    \hat{\vect{d}}^{v,\dev}_k = \dot{\gamma}^{v,\dev}_k \ \vect{N}^{\dev}_k
    \quad \text{where,} \quad \vect{N}^{\dev}_k = \frac{\vtau_{\neql(k)}^{\dev}}{||\vtau_{\neql(k)}^{\dev}||}
    \qquad \text{and} \qquad
    \hat{\vect{d}}^{v,\vol}_k = \frac{1}{3} \dot{\epsilon}^{v,\vol}_k  \mathds{1}.
\end{equation}
Here, $\dot{\gamma}^{v,\dev}_k$ and $\dot{\epsilon}^{v,\vol}_k$ are the effective deviatoric and volumetric viscous stretch rates, respectively, and $\vect{N}^{\dev}_k$ can be considered as the direction of deviatoric viscous flow with $||\vtau_{\neql(k)}^{\dev}|| = \sqrt{\vtau_{\neql(k)}^{\dev} : \vtau_{\neql(k)}^{\dev}}$ defined as the effective deviatoric Kirchhoff stress for the $k-$th viscous mechanism. Thus, we can express the evolution law for the $k-$th viscous mechanism in the following form:
\begin{equation}    \label{eq:stretch-rate-evolution-law}
    -\frac{1}{2} \mathcal{L}_v \vect{b}^e_k {\vect{b}^e_k}^{-1}
    = \hat{\vect{d}}^v_k 
    = \dot{\gamma}^{v,\dev}_k \frac{\vtau_{\neql(k)}^{\dev}}{||\vtau_{\neql(k)}^{\dev}||}
    + \frac{\dot{\epsilon}^{v,\vol}_k}{3}  \mathds{1}.
\end{equation}
To ensure the thermodynamic consistency of the residual dissipation Eq. \eqref{eq:residual-dissipation-split}, the constitutive descriptions for the effective viscous stretch rates need to ensure $\dot{\gamma}^{v,\dev}_k \geq 0$ and $p_{\neql(k)} \cdot \dot{\epsilon}^{v,\vol}_k \geq 0$. Due to the complexities associated with performing bulk relaxation experiments, it is challenging to develop insights and consequently prescribe a constitutive model for the volumetric viscous stretch rate behavior of amorphous elastomers. Hence, in most literature, viscous behavior is commonly assumed to be deviatoric. However, we decided to retain the volumetric viscous stretch rate term in Eq. \eqref{eq:stretch-rate-evolution-law} for the sake of retaining the generality of the theory.

\section{Constitutive model for elastomers exhibiting strain rate-induced glass transition}
\label{sec:constitutive-model}

\begin{figure}[htp]
    \centering
    \includegraphics[width=0.65\textwidth, page=2, clip, trim=8cm 7.75cm 8cm 8cm]{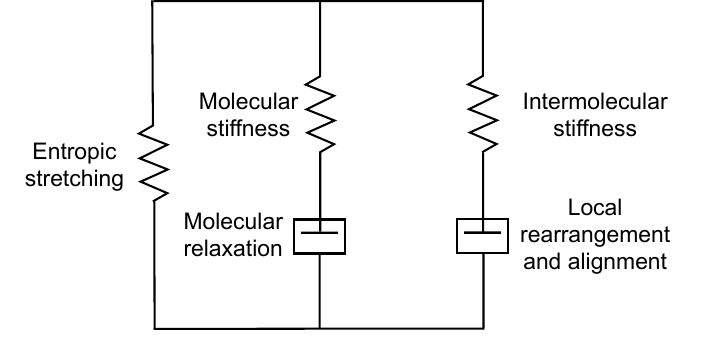}
    \caption{Rheological representation of dual-mechanism finite viscoelasticity model for quasi-static and high strain rate response of amorphous elastomers undergoing glass transition.}
    \label{fig:dual-mechanism-maxwell-model}
\end{figure}

In this section, we utilize the finite viscoelastic theory presented in Section \ref{sec:finite-viscoelasticity-theory} and specialize it with specific constitutive models to capture the rubbery and glassy behaviors of amorphous elastomers in quasi-static and high strain rate regimes, respectively. In particular, we assume two different viscous mechanisms, molecular relaxation at low strain rates and intermolecular rearrangement at high strain rates, are simultaneously active within the material, as illustrated in Figure \ref{fig:dual-mechanism-maxwell-model}. We start the section by prescribing the free energy densities for equilibrium and non-equilibrium processes, and then prescribe the constitutive equations for the viscous flow and the associated resistance corresponding to each mechanism under consideration. Following that, we describe the parameter identification procedure, then examine the quality of the model predictions for each strain‑rate regime, and briefly discuss the implications of the obtained material parameters.

\subsection{Free energy density functions and stress relations}

For amorphous elastomers, equilibrium response can be identified from the stress-strain response at a slow loading rate, which allows the polymer chains to remain in equilibrium between the loading states. Well-known Neo-Hookean and the Arruda-Boyce models \citep{arrudaThreedimensionalConstitutiveModel1993} are rooted in the statistical mechanics-based description of the entropic elasticity of the polymer chains; the latter being more generalized in its formulation with an additional parameter controlling the entropic resistance-induced stiffening. Hence, we chose the following form of the Arruda-Boyce hyperelastic potential \citep{arrudaThreedimensionalConstitutiveModel1993} to describe the equilibrium response of the polymer network:
\begin{equation} \label{eq:equilibrium-free-enery}
    \Psi_{\eq}^{\dev} (\barVect{C})  = G_{\eq} \lambda_{\mathrm{L}}^2 \left[ \left( \frac{\overline{\lambda}}{\lambda_{\mathrm{L}}} \right) \beta + \ln \left( \frac{\beta} {\sinh \beta} \right) \right]
    \ \text{and} \
    \Psi_{\eq}^{\vol}(J)             = \frac{\kappa_{\eq}}{4} (J^2 - 2 \ln J - 1)
\end{equation}
where $\overline{\lambda} = \sqrt{\frac{\tr(\barVect{C})}{3}}$ is the effective deviatoric stretch and $\beta = \mathcal{L}^{-1} \left(  \frac{\overline{\lambda}}{ \lambda_{\mathrm{L}}} \right)$, and $\mathcal{L}^{-1} (\bullet)$ is the inverse of the Langevin function defined as $\mathcal{L} (\bullet) = \coth(\bullet) - \frac{1}{(\bullet)}$. Material parameters, $G_{\eq}$, $\lambda_{\mathrm{L}}$, and $\kappa_{\eq}$, represent the equilibrium shear modulus, locking stretch, and equilibrium bulk modulus, respectively, for the entropic stretching of the elastomeric network. By substituting the prescribed equilibrium free energy density Eq. \eqref{eq:equilibrium-free-enery} in Eq. \eqref{eq:thermodynamic-restriction-stress-split}$_{1-2}$, we can obtain the deviatoric and volumetric parts of the equilibrium Cauchy stress as follows:
\begin{equation}    \label{eq:equilibrium-cauchy-stress}
    \vsigma_{\eq}^{\dev}        = J^{-1} \left[ \frac{G_{\eq}}{3} \frac{ \beta \lambda_{\mathrm{L}}}{\overline{\lambda}} \dev(\barVect{b}) \right]
    \quad \text{and} \quad
    \vsigma_{\eq}^{\vol}        = J^{-1} \left[ \frac{\kappa_{\eq}}{2} \left( J^2 - 1 \right) \mathds{1} \right],
\end{equation}
where $\dev(\barVect{b})$ represents the deviatoric part of the isochoric left Cauchy-Green tensor $\barVect{b} = \barVect{F} \barVect{F}^{\T} $.

\begin{remark}
    The Arruda-Boyce model reduces to the Neo-Hookean model in the limit of infinite locking stretch, \emph{i.e.}, $\lambda_{\mathrm{L}} \to \infty$. We should also note that, if the topology of the polymer network is complicated, where the statistical mechanics argument of the Arruda-Boyce model is no longer valid or the computation of the inverse Langevin function causes numerical singularity, it is also possible to adopt the Gent model \citep{gentNewConstitutiveRelation1996} to describe the entropic resistance-induced stiffening of the polymer in a phenomenological manner.
\end{remark}

Now, for each viscous mechanism, $k$, we assume a Neo-Hookean type potential with non-equilibrium shear moduli, $G_{\neql(k)}$, to represent the resistance to flow and a penalty function for the volumetric part characterized by the non-equilibrium bulk moduli, $\kappa_{\neql(k)}$, to represent the bulk deformation resistance as follows:
\begin{equation} \label{eq:non-equilibrium-free-energy}
    \Psi_{\neql(k)}^{\dev}(\vect{C}^e_k)    = \frac{G_{\neql(k)}}{2}  (\overline{I}_{1(k)}^e - 3) 
    \quad \text{and} \quad
    \Psi_{\neql(k)}^{\vol}(J^e_k)           = \frac{\kappa_{\neql(k)}}{4} ({J^e_k}^2 - 2 \ln J^e_k - 1),
\end{equation}
where $ \overline{I}_{1(k)}^e = \tr(\barVect{C}^e_k) $ is the first invariant of the isochoric elastic Cauchy-Green deformation tensor for the $k-$th viscous mechanism. Similar to the equilibrium stresses, the application of thermodynamic restrictions Eq. \eqref{eq:thermodynamic-restriction-stress-split}$_{3-4}$ on the non-equilibrium free energy density for any $k-$th viscous mechanism Eq. \eqref{eq:non-equilibrium-free-energy} yields the following non-equilibrium Cauchy stress components:
\begin{equation}    \label{eq:non-equilibrium-cauchy-stress}
    \vsigma_{\neql(k)}^{\dev}    =  J^{-1} \left[ G_{\neql(k)}  \dev(\barVect{b}^e_k) \right]
    \quad \text{and} \quad
    \vsigma_{\neql(k)}^{\vol}  = J^{-1} \left[ \frac{\kappa_{\neql(k)}}{2} \left( {J^e_k}^2 - 1 \right) \mathds{1} \right],
\end{equation}
where $\dev(\barVect{b}^e_k)$ represents the deviatoric part of the isochoric left Cauchy-Green tensor $\barVect{b}^e_k = \barVect{F}^e_k {\barVect{F}^e_k}^{\T} $. In earlier works on glassy polymers and rubbery elastomers, the authors either used the St. Venant-Kirchhoff potential \citep{boyceConstitutiveModelFinite2000,mullikenMechanicsRatedependentElastic2006,anandModelingMicroindentationResponse2006,amesThermomechanicallyCoupledTheory2009} or the Arruda-Boyce potential \citep{bergstromConstitutiveModelingLarge1998} to describe the resistance to viscous flow. From our numerical simulations, we found that the locking stretch parameter does not contribute to any additional hardening for either of the viscous mechanisms; however, it adds an additional parameter for each viscous mechanism to be determined, which may not be unique. Hence, we opted for a rather simple compressible Neo-Hookean type potential for both mechanisms.

\subsection{Viscous mechanism I: Molecular relaxation}

To complete the constitutive description, we now turn our attention to prescribing the effective viscous stretch rates, $\dot{\gamma}^{v,\dev}_{(k)}$, for viscous flows. When an elastomeric material, comprising a crosslinked network and free chain segments, is perturbed from its equilibrium state, Brownian motion of free and inactive chains along a constrained tube governs the time-dependent non-equilibrium behavior of elastomeric rubbers \citep{degennesReptationPolymerChain1971,doiTheoryPolymerDynamics1986}. Additionally, as the chain segments approach equilibrium, they often need to overcome a stress-driven energy barrier. Taking these micromechanisms into account, \cite{bergstromConstitutiveModelingLarge1998} proposed the following effective deviatoric viscous stretch rate formulation for rubber-like elastomers:
\begin{equation}   \label{eq:bergstrom-boyce-law}
    \dot{\gamma}^{v,\dev}_{(1)} = \gamma_{0} \left( \overline{\lambda}^v_{(1)} - 1 + \delta \right)^c \left( \frac{||\vtau_{\neql(1)}^{\dev}||} { \hat{\tau} } \right)^m,
\end{equation}
where $\gamma_0$, $c$, $\hat{\tau}$, and $m$ are the material parameters. We additionally introduced the stretch regularization constant, $\delta$, to avoid numerical singularity when viscous stretch, $\overline{\lambda}^v_{(1)} = \sqrt{ \tr \left( \barVect{C}^v_{(1)} \right)/3 } $, is small. In our numerical simulations, we set $\delta = 10^{-3}$, instead of calibrating it. By defining $ c_1 = \gamma_0/\hat{\tau}^m$, we can express the effective deviatoric viscous stretch rate Eq. \eqref{eq:bergstrom-boyce-law} as:
\begin{equation} \label{eq:bergstrom-boyce-law-redefined}
    \dot{\gamma}^{v,\dev}_{(1)} = c_1 \left( \overline{\lambda}^v_{(1)} - 1 + \delta \right)^{c_2} ||\vtau_{\neql(1)}^{\dev}||^m.
\end{equation}
The above form of the effective deviatoric viscous stretch rate, given in Eq. \eqref{eq:bergstrom-boyce-law-redefined}, is favorable as it reduces the number of material parameters to be identified and satisfies thermodynamic consistency Eq. \eqref{eq:residual-dissipation-split} for $c_1 \geq 0$ and $m \geq 0$, and based on the dynamics of polymer chains, $c_2 \in [-1, 0]$. Assuming the relation between the effective deviatoric stretch rate and effective deviatoric non-equilibrium stress as $\dot{\gamma}^{v,\dev}_{(1)} = \frac{1}{2 \nu_{(1)}^{\dev}} ||\vtau_{\neql(1)}^{\dev}||$, we can express the viscosity for the molecular relaxation mechanism, $\nu_{(1)}^{\dev}$, as:
\begin{equation}    \label{eq:bergstrom-boyce-viscosity}
    \nu_{(1)}^{\dev} = \frac{1}{2 c_1} \left( \overline{\lambda}^v_{(1)} - 1 + \delta \right)^{-c_2} ||\vtau_{\neql(1)}^{\dev}||^{1-m}.
\end{equation}

\subsection{Viscous mechanism II: Intermolecular rearrangement and alignment}

If the applied strain rate is above the threshold for glass transition to take place in an elastomer, the molecular relaxation with a large characteristic relaxation time is no longer a viable mechanism for viscous flow. Rather, viscous flow in the glassy state is attributed to the intermolecular resistance to rotation of macromolecular segments, which causes the polymer chains to be \enquote{frozen} in their configurations \citep{argonPlasticDeformationMetallic1979,boyceConstitutiveModelFinite2000}. As the polymer chains overcome the resistance, yielding occurs, and chain segments continue to align with the flow direction. Although \cite{anandModelingMicroindentationResponse2006} prescribed a phenomenological flow rule for amorphous glassy polymers, thus far, Ree-Eyring flow rule \citep{eyringViscosityPlasticityDiffusion1936,reeTheoryNonNewtonianFlow1955,reeTheoryNonNewtonianFlow1955a} turned out to be the most accurate way to describe the yielding across glass transition \citep{boyceLargeInelasticDeformation1988,mullikenMechanicsRatedependentElastic2006}. Hence, we adopted the following form of the Ree-Eyring flow rule \citep{nguyenThermoviscoelasticModelAmorphous2008} to describe the effective deviatoric viscous stretch rate, $\dot{\gamma}^{v,\dev}_{(2)}$, in the glassy state:
\begin{equation}    \label{eq:ree-eyring-law}
    \dot{\gamma}^{v,\dev}_{(2)} = \frac{\tau_y \theta}{\nu_0 Q_s} \exp \left( - \frac{\Delta G}{R \theta} \right) \sinh \left( \frac{Q_s  ||\vtau_{\neql(2)}^{\dev}|| }{\tau_y \theta} \right),
\end{equation}
where $R$ denotes the universal gas constant, $\theta$ is the absolute temperature, $\tau_y$ is the yield stress, $\nu_0$ is the reference viscosity, $Q_s$ is the stress activation energy, and $\Delta G$ is the thermal activation energy. Amorphous glassy polymers, \emph{i.e.}, elastomers deformed at high strain rates beyond the threshold, show significant strain hardening, causing the yield surface to evolve, which can be described using the following evolution law for the yield stress:
\begin{equation}    \label{eq:yield-stress-evolution}
    \dot{\tau}_y = h \left( 1 + \frac{\tau_y}{\tau_y^0} \right) \dot{\gamma}^{v,\dev}_{(2)},
\end{equation}
where $\tau_y^{0}$ is the initial yield strength and the hardening modulus, $h$, determines the rate of strain hardening. Viscous flow in amorphous glassy polymers, such as polycarbonate (PC) and poly(methyl methacrylate) (PMMA), exhibits strong dependency on the temperature and pressure with strain softening \citep{boyceLargeInelasticDeformation1988,boyceConstitutiveModelFinite2000,mullikenMechanicsRatedependentElastic2006}. If an elastomer exhibits such characteristics at high strain rates, the yield stress evolution Eq. \eqref{eq:yield-stress-evolution} can be modified to incorporate the necessary effects. Finally, we should note that since the absolute temperature, $\theta$, and the effective non-equilibrium deviatoric Kirchhoff stress, $||\vtau_{\neql(2)}^{\dev}||$, are always positive by definition, for the Ree-Eyring flow rule Eq. \eqref{eq:ree-eyring-law} to satisfy the positive dissipation criterion Eq. \eqref{eq:residual-dissipation-split}, all material parameters characterizing the process of intermolecular rearrangement and alignment need to be positive. Similar to the viscosity expression for the molecular relaxation mechanism Eq. \eqref{eq:bergstrom-boyce-viscosity}, we can express the viscosity for the intermolecular rearrangement and alignment, $\nu_{(2)}^{\dev}$, as:
\begin{equation}    \label{eq:ree-eyring-viscosity}
    \nu_{(2)}^{\dev} = \frac{\nu_0 Q_s}{2 \tau_y \theta} \exp \left(\frac{\Delta G}{R \theta} \right) \left[ \sinh \left( \frac{Q_s  ||\vtau_{\neql(2)}^{\dev}|| }{\tau_y \theta} \right) \right]^{-1} ||\vtau_{\neql(2)}^{\dev}||.
\end{equation}

\subsection{Volumetric viscosity and relaxation}

The last remaining component of our model to be prescribed is the constitutive law for the volumetric viscous stretch rate, $\dot{\epsilon}^{v,\vol}_k$, corresponding to the $k-$th viscous mechanism. As mentioned earlier, due to the lack of experimental observations in current literature, we decided to assume the simplest possible constitutive law for volumetric viscous stretch rate, $\dot{\epsilon}^{v,\vol}_k$, given below:
\begin{equation}    \label{eq:volumetric-creep-law}
    \dot{\epsilon}^{v,\vol}_k = \frac{p_{\neql(k)}}{\nu^{\vol}_k}
    \qquad \text{with} \ k \in \{1,2\},
\end{equation}
where $p_{\neql(k)} = \frac{1}{3} \tr (\vtau_{\neql}^{\vol})$ is known as the non-equilibrium pressure and $\nu^{\vol}_{k}$ is referred to as the volumetric viscosity related to the $k-$th viscous mechanism in the material. The positive dissipation criterion Eq. \eqref{eq:residual-dissipation-split} requires the volumetric viscosity to be positive, \emph{i.e.}, $\nu^{\vol}_{k} > 0$, for the prescribed constitutive Eq. \eqref{eq:volumetric-creep-law} to be thermodynamically-consistent.
\begin{remark}
    With the expressions for the deviatoric viscosities, $\nu_{(1)}^{\dev}$ and $\nu_{(2)}^{\dev}$, given in Eq. \eqref{eq:bergstrom-boyce-viscosity} and Eq. \eqref{eq:ree-eyring-viscosity}, respectively, and the volumetric viscosities, $\nu^{\vol}_{(1)}$ and $\nu^{\vol}_{(2)}$, given in Eq. \eqref{eq:volumetric-creep-law}, we can relate our proposed constitutive model to the fourth-order viscosity tensor Eq. \eqref{eq:RG-inverse-viscosity-tensor} proposed by \cite{reeseTheoryFiniteViscoelasticity1998}.
\end{remark}

\subsection{Identification of material parameters}

We implemented our material model in MATLAB using a predictor-corrector-based explicit time integration scheme, which closely follows the implicit finite element implementation algorithm proposed by \cite{reeseTheoryFiniteViscoelasticity1998}. The details of the constitutive computation procedure can be found in \ref{sec:numerical-implementation-procedure}. In \cite{konaleLargeDeformationModel2023}, the authors reported stress-strain response of polyborosiloxane (PBS), an elastomer with reversible crosslinks, under quasi-static compression ($\dot{\varepsilon} = 0.001-1$ s$^{-1}$) and split Hopkinson pressure bar (SHPB) tests ($\dot{\varepsilon} = 4500$ s$^{-1}$ and $8500$ s$^{-1}$). We used our material model in conjunction with uniaxial stress boundary condition, \emph{i.e.}, $\sigma_{22} = \sigma_{33} = 0$, to identify the material parameters by comparing experimentally reported true stress, $\sigma_{11}$, with the applied strain, $\varepsilon_{11}$, using a nonlinear least-squares optimizer, \texttt{lsqnonlin}, from MATLAB. As detailed below, we employed a three-step procedure to identify the material parameters of our proposed constitutive model based on the experimental data from \cite{konaleLargeDeformationModel2023}:

%%%% calibration figure
\begin{figure}[htp]
    \centering
    \includegraphics[width=\textwidth, page=3, clip, trim=2.25cm 6cm 1.8cm 6cm]{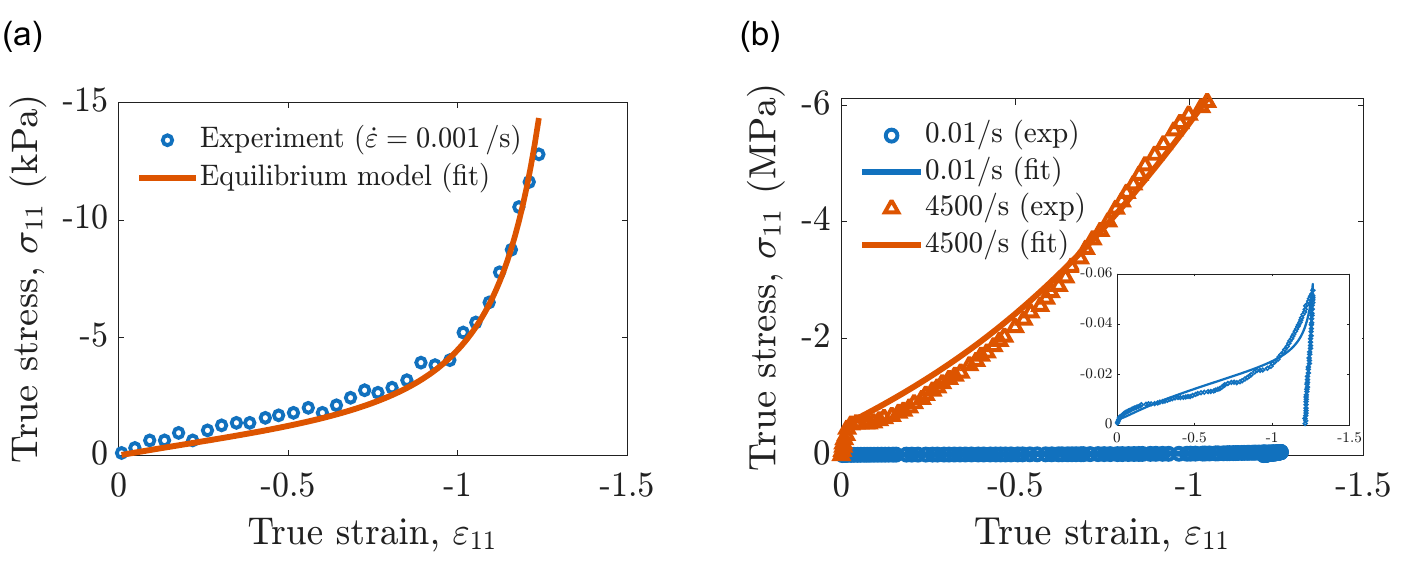}
    \caption{Identification of material parameters: (a) calibration of the equilibrium shear modulus, $G_{\eq}$, and locking stretch, $\lambda_{\mathrm{L}}$, from lowest strain rate experimental data available ($\dot{\varepsilon} = 10^{-3}$ s$^{-1}$), with a coefficient of determination, $R^2 = 0.9780$, (b) calibration of Bergstr\"{o}m-Boyce and Ree-Eyring parameters from compressive loading-unloading experiment performed at a strain rate, $\dot{\varepsilon} = 10^{-2}$ s$^{-1}$, with a coefficient of determination, $R^2 = 0.9593$, and split Hopkinson pressure bar (SHPB) test performed at a strain rate, $\dot{\varepsilon} = 4500$ s$^{-1}$, with a coefficient of determination, $R^2=0.9878$, respectively. Inset shows model fit at $\dot{\varepsilon} = 10^{-2}$ s$^{-1}$. All experimental data were taken from \cite{konaleLargeDeformationModel2023}}.
    \label{fig:konale-data-fit}
\end{figure}

\begin{itemize}[leftmargin=0.45cm, itemsep=0pt, topsep=-3pt]

    \item \textbf{Step 1:} We first used the compressive stress-strain data reported at a strain rate of $10^{-3}$ s$^{-1}$ to determine the equilibrium material parameters: equilibrium shear modulus, $G_{\eq}$, and locking stretch, $\lambda_{\mathrm{L}}$. Since the material is in its rubbery state at this strain rate, we set the bulk modulus, $\kappa_{\eq} = 50 G_{\eq} $, to enforce the near-incompressible behavior of the material. In this step, we ignored both viscous mechanisms and obtained the parameters, $G_{\eq} = 622.04$ Pa and $\lambda_{\mathrm{L}} = 1.58$, with a coefficient of determination, $R^2 = 0.9780$, using the loading data, as shown in Figure \ref{fig:konale-data-fit}(a). 

    \item \textbf{Step 2:} In this step, we used the loading and unloading data at a strain rate, $\dot{\varepsilon} = 10^{-2}$ s$^{-1}$, to determine the following parameters related to the molecular relaxation mechanism: non-equilibrium shear modulus, $G_{\neql(1)}$, stretch pre-factor, $c_1$, stretch exponent, $c_2$, and stress exponent, $m$. Following the same reasoning stated in Step 1, we set the non-equilibrium bulk modulus, $\kappa_{\neql(1)} = 50 G_{\neql(1)}$. Besides equilibrium, we kept both viscous mechanisms (the complete form of our constitutive model) active in this step; however, the small characteristic relaxation time of intermolecular rearrangement and alignment ensures the viscous strain contribution from intermolecular flow is negligible at this strain rate. Recognizing this fact, we set $G_{\neql(2)}$ to be of the same order of magnitude as $G_{\neql(1)}$, which allowed us to use a larger time step, $\Delta t = 10^{-2}$ s, in turn accelerating the optimization process. Through an iterative optimization process, we obtained $G_{\neql(1)} = 226.4$ kPa, $m=2.7$, $c_1 = 250$ s$^{-1}$ MPa$^{-m}$, and $c_2 = -1$ with a reported coefficient of determination, $R^2 = 0.9593$. A comparative stress-strain plot is shown in Figure \ref{fig:konale-data-fit}(b) with the obtained parameters and the experimentally reported data.

    \item \textbf{Step 3:} In this step, we first determined the non-equilibrium shear modulus, $G_{\neql(2)}$, from the initial stress-strain data at the strain rate of 4500 s$^{-1}$. We assumed the non-equilibrium bulk modulus, $\kappa_{\neql(2)} = 5 G_{\neql(2)}$, to ensure sufficient compressibility of the glassy state at high strain rate. We then estimated the reference viscosity, $\nu_0 = \frac{G'^2 + G''^2}{\omega_0 G''} = 3.53 \times 10^{3}$ Pa$\cdot$s, where $G' = 33270.46$ and $G'' = 3162.28$ at $\omega_0 = 100$ rad/s from oscillatory shear data reported by the authors. In their works, \cite{boyceConstitutiveModelFinite2000,mullikenMechanicsRatedependentElastic2006,nguyenThermoviscoelasticModelAmorphous2008} reported the thermal activation energy, $\Delta G$, for different glassy amorphous polymers to be in the range of $10^2-10^5$ J/mol. Due to the lack of experimental data at different temperatures, we set $\Delta G = 10^3$ J/mol for this material. We retained all the parameters obtained from the first two steps and continued to optimize the following parameters: initial yield strength, $\tau_y^0$, hardening modulus, $h$, and stress activation energy, $Q_s$, related to intermolecular flow by comparing against stress-strain data at the strain rate, $\dot{\varepsilon} = 4500$ s$^{-1}$. From the iterative optimization process with time step $\Delta t = 10^{-7}$ s, we obtained $\tau_y^0 = 11.183$ MPa, $h = 25.24$ MPa, and $Q_s = 5 \times 10^4$ K with a coefficient of determination, $R^2 = 0.9878$, and plotted the fitted stress-strain response against the experimental data in Figure \ref{fig:konale-data-fit}(b). 
    
\end{itemize}

The equilibrium shear modulus, $G_{\eq} = 622.04$ Pa, is comparable to values reported for putty-like soft elastomeric materials, confirming that PBS is highly compliant in the relaxed state. In contrast to the low equilibrium shear modulus, $G_{\eq}$, the non-equilibrium shear modulus in the rubbery-state, $G_{\neql(1)}$, is approximately 360 times larger in magnitude, which is a consequence of the large residual strain exhibited by the material upon unloading. Although the locking stretch, $\lambda_{\mathrm{L}}$, was determined as an equilibrium state parameter, the hardening observed at low strain rates is a consequence of the \emph{back stress} induced by the resistance to entropic stretching. The non-equilibrium shear modulus for the glassy state, $G_{\neql(2)} = 9.72$ MPa, represents the extreme rate-stiffening at high strain rate, which is a hallmark of \enquote{frozen chain} configuration. It is ideal to estimate the reference viscosity, $\nu_0$, at a frequency where the glass transition has already occurred and possibly close to the strain rate where the rest of the parameters corresponding to the second mechanism are to be calibrated (\textbf{Step 3} as described above). In this case, we used the data at the reported maximum angular frequency (100 rad/s) despite being about an order of magnitude lower than our strain rate of interest. A complete list of the material parameters obtained from our multi-step parameter identification procedure is given in Table \ref{tab:PBS_konale_calibrated properties}; these parameters were then used in performing numerical simulations presented in Section \ref{sec:results-discussion}.

\begin{table}[!ht]
\centering
\begin{tabular}{p{2cm} c c c c}
    \toprule \toprule
    \textbf{Mechanism}  & \textbf{Material parameters}    & \textbf{Symbol}       & \textbf{Value}  & \textbf{Unit}     \\
    \toprule \toprule
    \multirow{3}{*}{\makecell{Entropic \\ elasticity}}
    & Equilibrium shear modulus         & $G_{\eq}$                 & $622.04$       & Pa          \\
    & Locking stretch                   & $\lambda_{\text{L}}$      & $1.58$        & $-$         \\
    & Equilibrium bulk modulus          & $\kappa_{\eq}$            & $50G_{\eq}$   & Pa          \\
    \midrule
    \multirow{7}{*}{\makecell{Molecular \\ relaxation}}
    & Non-equilibrium shear modulus     & $G_{\neql(1)}$            & $226.4$       & kPa         \\
    & Non-equilibrium bulk modulus      & $\kappa_{\neql(1)}$       & $50G_{\neql(1)}$  & kPa     \\
    & Stretch pre-factor                & $c_1$                     & $250$         & MPa$^{-2.7}/$s    \\
    & Stretch exponent                  & $c_2$                     & $-1$          & $-$         \\
    & Stress exponent                   & $m$                       & $2.7$         & $-$         \\
    & Stretch regularization constant   & $\delta$                  & $10^{-3}$     & $-$         \\
    & Volumetric viscosity              & $\nu^{\vol}_{(1)}$        & $10^{15}$     & Pa$\cdot$s  \\
    \midrule
    \multirow{8}{*}{\makecell{Local \\ rearrangement \\ and \\ alignment}}
    & Non-equilibrium shear modulus     & $G_{\neql(2)}$            & $9.72$        & MPa         \\
    & Non-equilibrium bulk modulus      & $\kappa_{\neql(2)}$       & $5G_{\neql(2)}$ & MPa       \\   
    & Reference viscosity               & $\nu_0$                   & $3.53$        & kPa$\cdot$s \\
    & Initial shear strength            & $\tau_y^0$                & $11.183$      & MPa         \\
    & Hardening modulus                 & $h$                       & $25.24$       & MPa         \\
    & Stress activation energy          & $Q_s$                     & $5\times10^4$ & K    \\
    & Thermal activation energy         & $\Delta G$                & $1000$        & J/mol       \\
    & Volumetric viscosity              & $\nu^{\vol}_{(2)}$        & $10^{15}$     & Pa$\cdot$s  \\
    \bottomrule
\end{tabular}
\caption{List of material properties for polyborosiloxane (PBS) obtained from quasi-static and high strain rate experimental tests performed by \cite{konaleLargeDeformationModel2023}. Some of the material properties were set to reasonable values due to the lack of experimental data.}
\label{tab:PBS_konale_calibrated properties}
\end{table}

\section{Results and discussion with numerical examples}
\label{sec:results-discussion}

In this section, we demonstrate the essential features of our proposed constitutive model through numerical examples using the material parameters listed in Table \ref{tab:PBS_konale_calibrated properties} for polyborosiloxane (PBS). We numerically studied the degree of strain rate sensitivity and energy dissipation in a single and multiple loading-unloading cycles across a wide range of strain rates for large deformation, stress relaxation behavior of the material for large strains applied at different strain rates, and finally, viscoelastic characteristics of the material in the small strain regime. In what follows, we describe our numerical examples in detail and discuss the characteristics of our proposed constitutive model in detail, followed by the significance of the proposed constitutive model.

\subsection{Strain rate sensitivity under uniaxial compression}

%% strain rate study
\begin{figure}[!ht]
\centering
    \includegraphics[width=\textwidth, page=4, clip, trim=2.1cm 4.5cm 1.9cm 5.5cm]{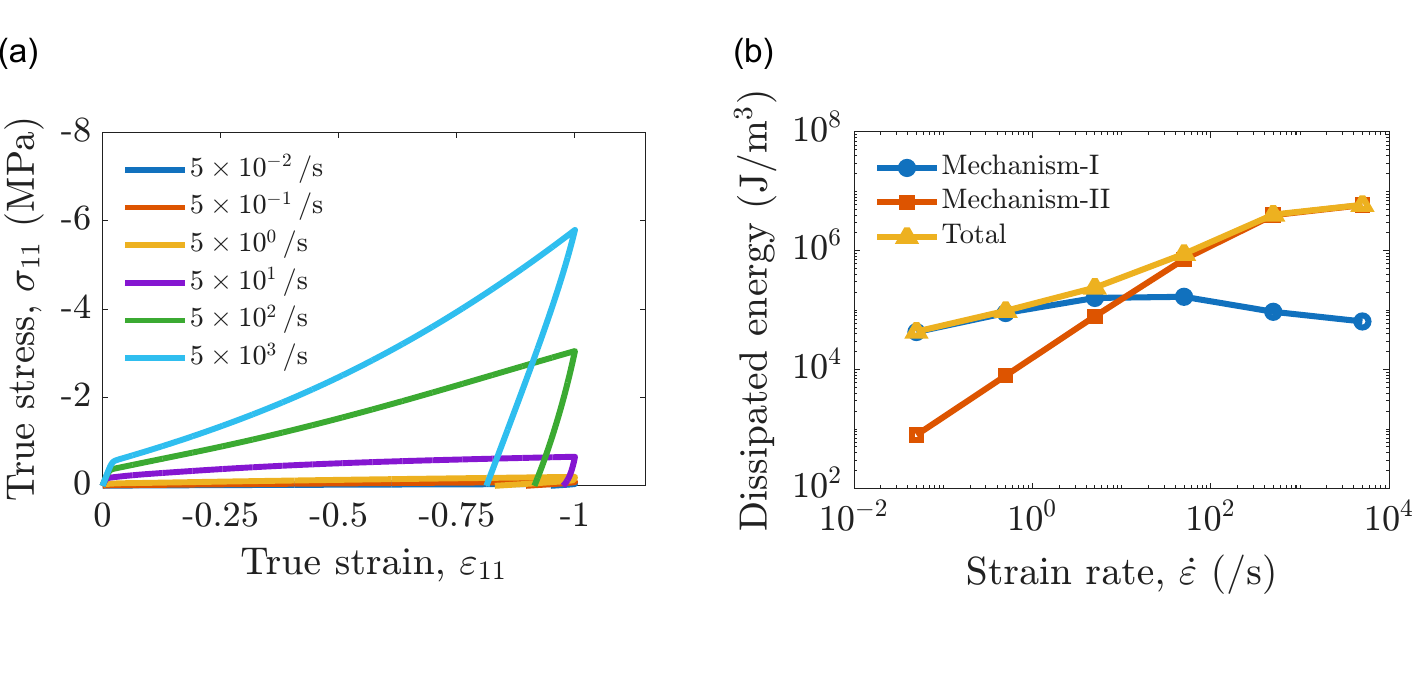}
    
    \vspace{-\baselineskip}
    
    \includegraphics[width=\textwidth, page=5, clip, trim=2.1cm 5.25cm 1.9cm 5.5cm]{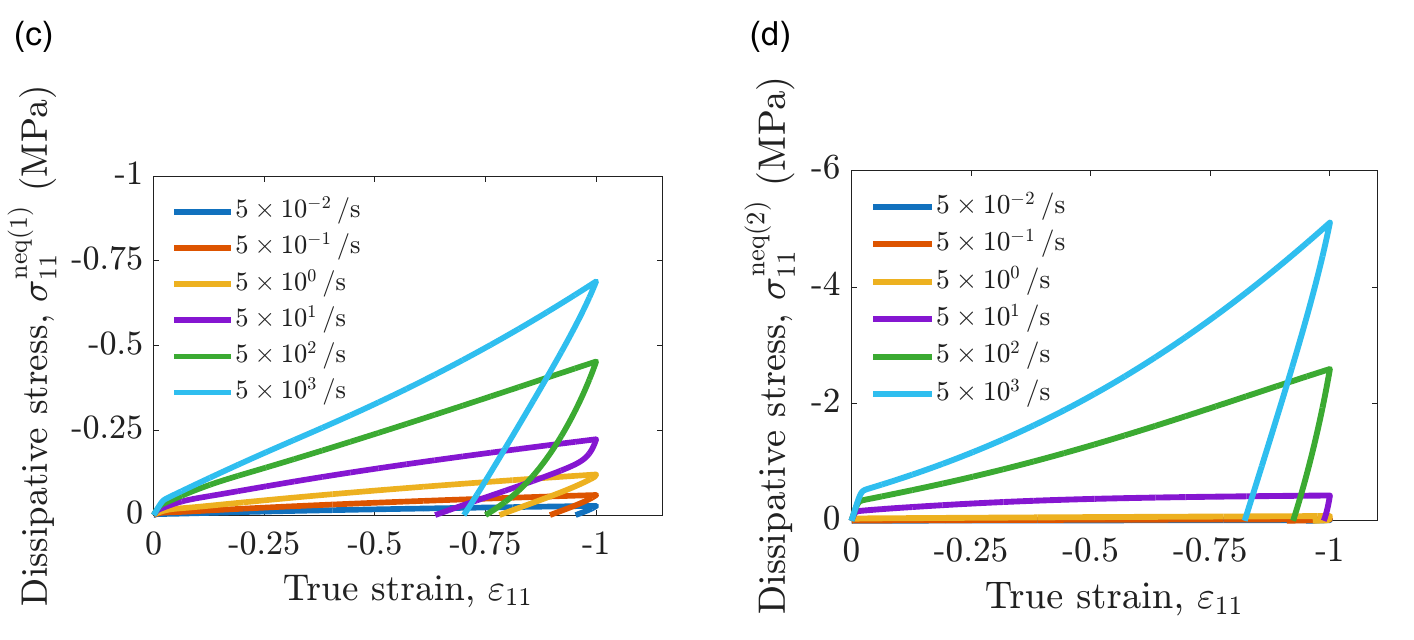}
\caption{Rate sensitivity study of the material model in uniaxial compressive loading-unloading: (a) stress-strain curve at different strain rates, (b) accumulated individual and total dissipation at different strain rates, contribution of non-equilibrium (dissipative) stress from (c) mechanism-I and (d) mechanism-II at different strain rates.}
\label{fig:strain-rate-example}
\end{figure}

In our first example, we performed a loading-unloading simulation of uniaxial compression up to 100\% true strain at true strain rates ranging from $5 \times 10^{-2}$ s$^{-1}$ to $5 \times 10^{3}$ s$^{-1}$. For the true strain rates, $5 \times 10^{-2}$ s$^{-1}$ to $5 \times 10^{1}$ s$^{-1}$, we set the time step, $\Delta t = 10^{-5}$ s, whereas for the true strain rates, $5 \times 10^{2}$ s$^{-1}$ and $5 \times 10^{3}$ s$^{-1}$, we set the time step, $\Delta t = 10^{-6}$ s and $10^{-7}$ s, respectively. For each strain rate case, at each time step, given the applied true strain, $\varepsilon_{11} = \ln (\lambda_1)$, where $\lambda_1$ is one of the principal stretch components, we solved for the principal stretch component, $\lambda_2$, by enforcing the boundary condition $\sigma_{22} = 0$. Because of the isotropic nature of our model, we set $\lambda_3 = \lambda_2$ to form the deformation gradient, $\vect{F}$, and in turn compute the true stress, $\sigma_{11}$, following the procedure given in \ref{sec:numerical-implementation-procedure}. We plotted the total true stress, $\sigma_{11}$, against the applied true strain, $\varepsilon_{11}$, in Figure \ref{fig:strain-rate-example}(a) for each strain rate. We can observe a significant increase in the peak stress value at the end of the compressive loading, almost by two orders of magnitude, as the applied strain rate was increased from the lowest to the highest. For the applied strain rates ranging from 50 s$^{-1}$ to 5000 s$^{-1}$, we can observe a stiffer initial response followed by yielding and strain hardening, which can be attributed to the local rearrangement and alignment of so-called \enquote{frozen} chains at high strain rates. To understand the role of viscous mechanisms, we plotted the non-equilibrium (dissipative) stress components, $\sigma_{11}^{\neql(1)}$ and $\sigma_{11}^{\neql(2)}$, pertaining to the individual mechanisms in Figure \ref{fig:strain-rate-example}(c)--(d). Based on the magnitude of the non-equilibrium stresses at different strain rates, it is evident that at low strain rates, the molecular relaxation is primarily active, and at high strain rates, the local rearrangement and alignment of the molecules become more active. This can be further confirmed from the dissipated energy density plot (see Figure \ref{fig:strain-rate-example}(b)) which shows dissipated energy attributed to the first mechanism (molecular relaxation) decreases beyond the strain rate of 0.5 s$^{-1}$ whereas the dissipated energy contribution attributed to the second mechanism (intermolecular rearrangement) increases monotonically with increasing strain rates. For the first mechanism, as the strain rate becomes higher, the dissipative stress at complete unloading becomes tensile, thereby making the dissipation over the loading-unloading cycle smaller. However, for the second mechanism, as the peak stress increases with increasing strain rates, the dissipation over the cycle is always increasing, eventually becoming the dominant one among the two underlying mechanisms. The crossover point for the dissipation mechanisms can be identified around the strain rate of 10 s$^{-1}$, where the intermolecular rearrangement and alignment start to become dominant over the molecular relaxation, indicating the glass transition is taking place. Based on the maximum total dissipated energy density, $W_{\mathrm{dis}}^{\mathrm{max}} = 5.94$ MJ/m$^3$, at the highest applied strain rate (5000 s$^{-1}$), we can estimate the temperature rise in the material to be 5 \dC, which eliminates the possibility of thermal softening, typically observed in many glassy materials\footnotemark. Since it is difficult to perform loading-unloading experiments at high strain rates, our model primarily relies on the unloading characteristics originating from the molecular relaxation mechanism because of the parameter identification procedure we followed in this work.

\footnotetext{The temperature rise, $\Delta T$, can be approximated using $\Delta T = \frac{W_{\mathrm{dis}}}{\rho C_p}$ where $W_{\mathrm{dis}}$ is the total dissipated energy turning into heat, and $\rho$ and $C_p$ represent the density and the isobaric specific heat capacity, respectively, of the material. For PBS, we assumed $\rho = 1050$ kg/m$^3$ and isobaric specific heat, $C_p = 1200$ J/kg-K, to compute the temperature change. }

\subsection{Cyclic loading-unloading}

%% cyclic loading unloading study
\begin{figure}[!ht]
\centering
    \includegraphics[width=\textwidth, page=6, clip, trim=2.2cm 5.2cm 1.8cm 5.5cm]{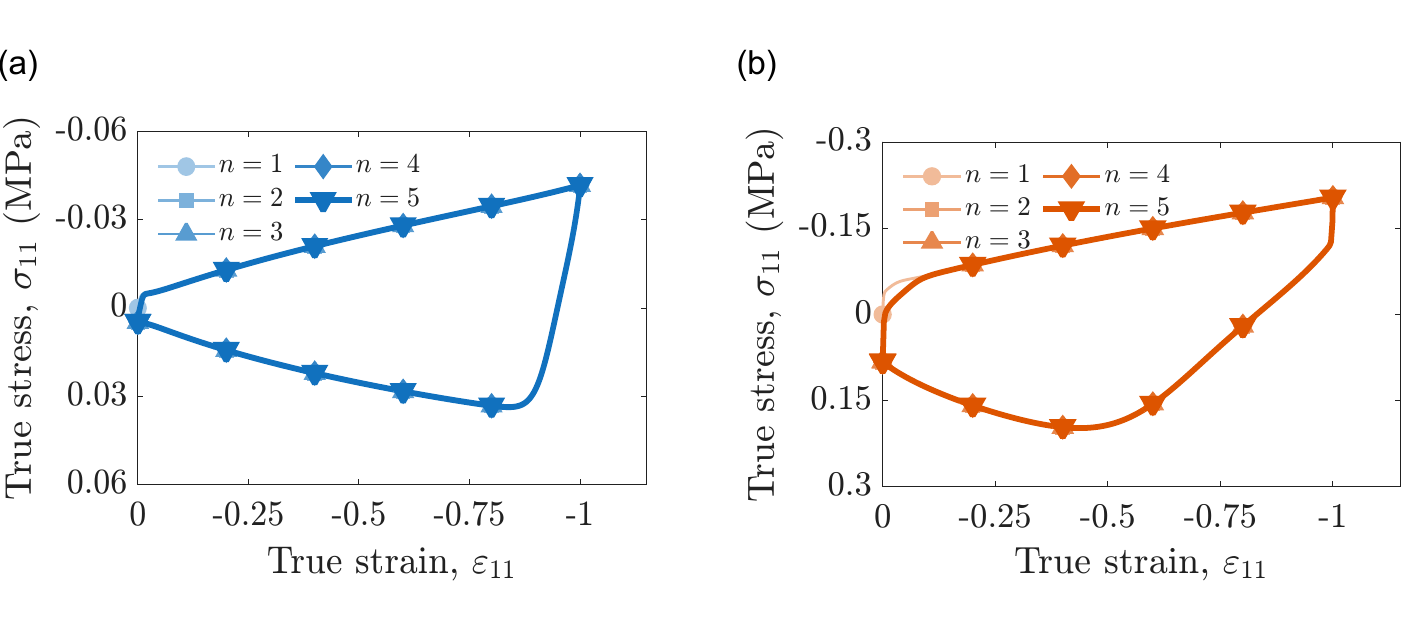}

    \vspace{-\baselineskip}
    
    \includegraphics[width=\textwidth, page=7, clip, trim=2.2cm 5.75cm 2.1cm 5.5cm]{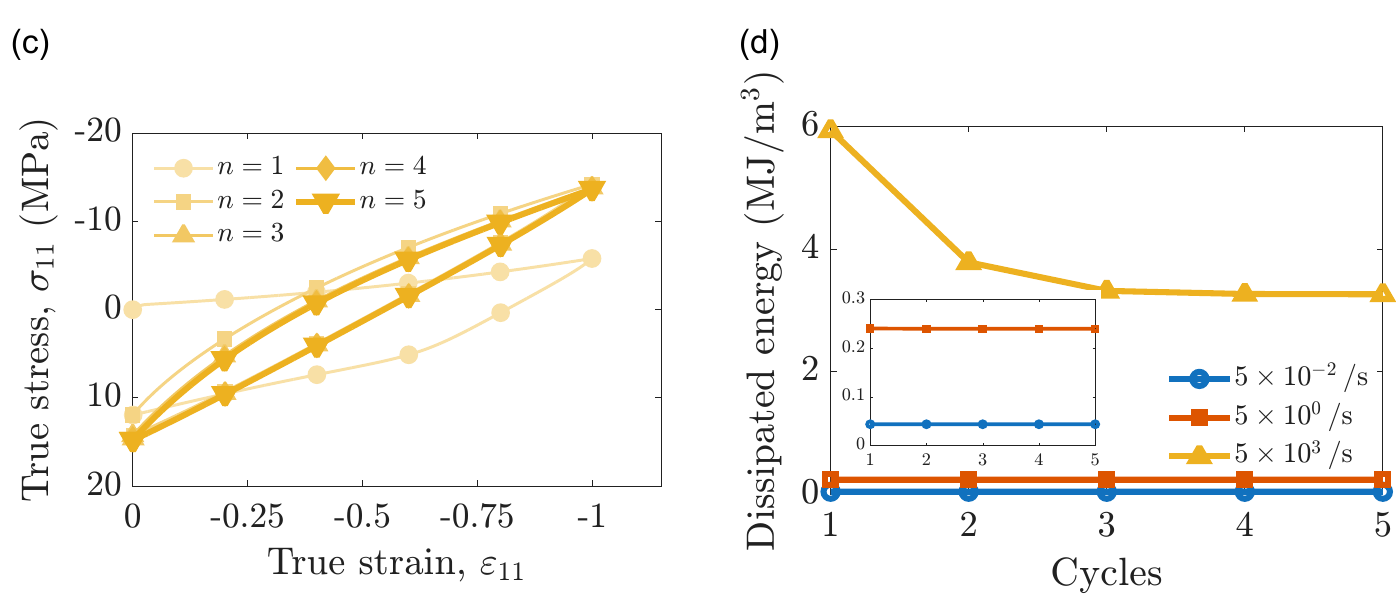}
\caption{Stress-strain response under cyclic compressive loading and unloading performed at (a) low ($\dot{\varepsilon} = 5 \times 10^{-2}$ s$^{-1}$), (b) moderate ($\dot{\varepsilon} = 5 $ s$^{-1}$), and (c) high ($\dot{\varepsilon} = 5 \times 10^3$ s$^{-1}$) strain rates where $n$ represents the cycle number, and (d) total dissipated energy density per cycle for different strain rates with an inset illustrating the dissipated energy at low and moderate strain rates.}
\label{fig:cyclic-loading-example}

\end{figure}

In this example, we extended the compressive loading-unloading condition applied in the previous example from a single cycle to multiple cycles to understand how the stress evolves under cyclic conditions. We kept the maximum applied true strain, material properties, and time step size corresponding to each strain rate the same as in the previous example, and performed the simulations at three representative true strain rates: $5 \times 10^{-2}$ s$^{-1}$ (low), $5$ s$^{-1}$ (moderate), and $5 \times 10^{3}$ s$^{-1}$ (high), for 5 cycles. We followed the same solution and constitutive computation procedure as the previous example. The resulting stress-strain response for the first to fifth cycles for all the applied strain rates is provided in Figure \ref{fig:cyclic-loading-example}(a)--(c). At the lowest true strain rate (0.05 s$^{-1}$), the non-equilibrium response is dominated by the molecular relaxation mechanism, which results in nearly identical stress-strain response from cycle to cycle (Figure \ref{fig:cyclic-loading-example}(a)). This can be further confirmed from Figure \ref{fig:cyclic-loading-example}(d), which shows the dissipative energy density to remain almost the same from cycle to cycle. On the other hand, at an intermediate true strain rate (5 s$^{-1}$), we can see that the magnitude of the dissipative energy density becomes stabilized after two cycles, and at the highest true strain rate (5000 s$^{-1}$), it takes about three cycles before the dissipative energy density becomes constant. We should note that the apparent drop in the dissipative energy density from cycle to cycle is larger at 5000 s$^{-1}$ than at 5 s$^{-1}$, as a larger fraction of the polymer network aligns with the flow direction at the higher strain rate, resulting in lower energy dissipation in subsequent cycles.. For both cases, we can see a decrease in the magnitude of the dissipative energy density, which is also evident from the change in size of the hysteresis loop over multiple cycles as depicted in Figure \ref{fig:cyclic-loading-example}(b)--(c), respectively. With the increasing number of loading-unloading cycles at high strain rate, as a higher volume fraction of the polymer network becomes aligned with the direction of the flow, the polymer network can carry more load elastically, causing a higher peak stress and a decrease in dissipative energy density.

\subsection{Stress relaxation}

\begin{figure}[!ht]
    \centering
    \includegraphics[width=\textwidth, page=8, clip, trim=0.8cm 7cm 0.8cm 6.75cm]{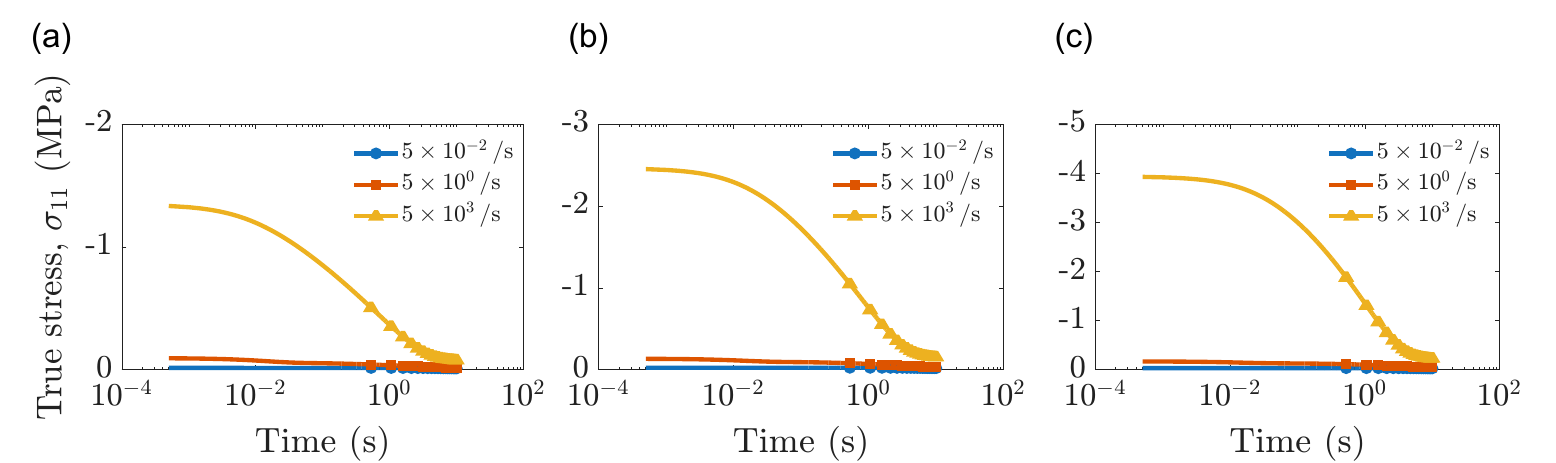}
    \caption{Stress relaxation characteristics of the material for the applied true strain of (a) 25\%, (b) 50\%, and (c) 75\%, at the applied true strain rates of: $\dot{\varepsilon} = 5 \times 10^{-2}$ s$^{-1}$ (low), $\dot{\varepsilon} = 5 $ s$^{-1}$ (moderate), and $\dot{\varepsilon} = 5 \times 10^{3}$ s$^{-1}$ (high) in the loading period.}
    \label{fig:relaxation-example}
\end{figure}

We now use numerical simulation to understand the stress relaxation behavior of PBS using the constitutive model described in this work. We prescribed true compressive strains of 25\%, 50\%, and 75\% at three representative true strain rates, $5 \times 10^{-2}$ s$^{-1}$, $5$ s$^{-1}$, and $5 \times 10^{3}$ s$^{-1}$. We then held the applied true strain for a time period of 10 s to allow stress relaxation. During loading, we kept the time step, $\Delta t$, the same as the previous examples; however, during the hold period, we increased it to be $5 \times 10^{-4}$ s to speed up the simulations. We plotted the stress relaxation curves for each applied strain case in Figure \ref{fig:relaxation-example}(a)--(c). As expected, the peak stress at the end of the loading period is highest for the maximum strain rate case for all three prescribed strain levels. This is because at the highest loading rate, due to the lack of time, viscous mechanisms do not \enquote{kick in} to relax the material and a larger stress develops at the end of the loading period. However, as soon as the loading process ends and the material is held at the maximum prescribed strain, viscous flow begins. Because of the inherent nonlinear nature of the prescribed deviatoric viscosities in our constitutive model, as given in Eq. \eqref{eq:bergstrom-boyce-viscosity} and Eq. \eqref{eq:ree-eyring-viscosity}, we can see that the maximum stress at the end of the loading period does not scale linearly with the level of applied strain for the same strain rate. For all three levels of the prescribed strain, the highest strain rate cases have the largest peak stress, which is driving the viscous flow at the beginning of the hold period. With time, the dramatic decrease in stress slows down the viscous flow and consequently the material reaches the equilibrium state. On the other hand, at the lowest loading rate, the material has sufficient time to undergo molecular relaxation as the loading process is ongoing. Hence, the maximum stress is significantly smaller compared to the highest strain rate case, and relaxation is almost instantaneous. 

%% stress relaxation study

\subsection{Small amplitude oscillatory tension-compression}

\begin{figure}[!ht]
    \centering
    \includegraphics[width=0.7\textwidth, page=9, clip, trim=7cm 6.75cm 7cm 6cm]{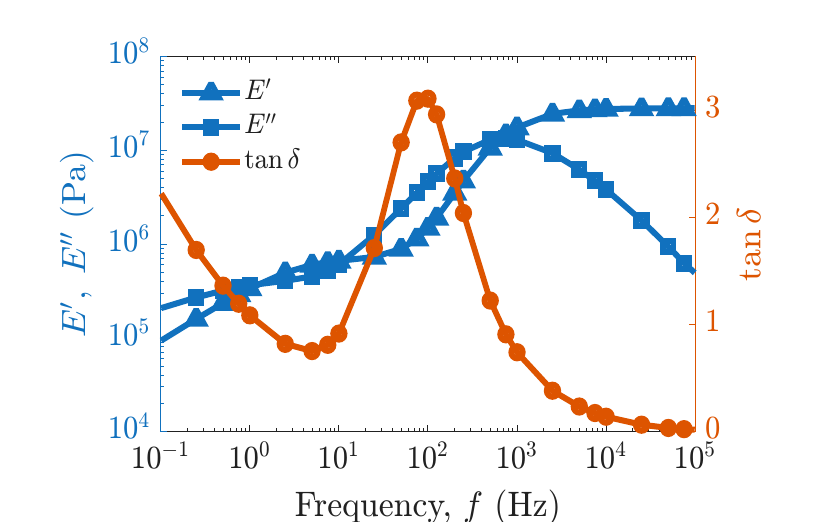}
    \caption{Variation of the storage modulus, $E^{\prime}$, loss modulus, $E^{\prime\prime}$, and loss factor, $\tan \delta$, from 0.1 Hz to $10^5$ Hz for a small amplitude ($\varepsilon_0$ = 0.01) oscillatory tensile-compressive loading.}
    \label{fig:oscillation-example}
\end{figure}

In our final example, we numerically studied the linear viscoelastic behaviors of the material with our proposed model in the small-strain regime. We prescribed uniaxial tension-compression loading following the strain profile, $\varepsilon(t) = \varepsilon_0 \sin (2\pi ft)$, where $\varepsilon_0$ is the strain amplitude and $f$ is the oscillation frequency. We chose the strain amplitude, $\varepsilon_0 = 0.01$, to ensure the deformation is within the linear viscoelastic regime, and varied the frequency from 0.1 Hz to $10^5$ Hz. For frequencies in the range of 0.1 Hz to 1000 Hz, we used a time step, $\Delta t = 10^{-5}$ s, and for frequencies above 1000 Hz, we used a time step, $\Delta t = 10^{-6}$ s. Application of the uniaxial stress boundary condition and further constitutive computation are the same as described in the previous examples. The prescribed oscillatory loading condition was continued for 5 cycles at each frequency to ensure the viscous dissipation is stabilized. We then used the stress-strain response from the last cycle at all frequencies to perform a linear least-squares procedure to compute the storage modulus, $E^{\prime}$, the loss modulus, $E^{\prime\prime}$, and the loss factor, $\tan \delta$, \citep{ferryViscoelasticPropertiesPolymer1980}, which are illustrated in Figure \ref{fig:oscillation-example}.

%% dynamic mechanical analysis study

The computed storage modulus, $E^{\prime}$, increases from one equilibrium to another equilibrium stage undergoing glass transition. At the limit of lowest frequency (0.1 Hz), the storage modulus, $E^{\prime}$, corresponds to the rubbery modulus, which is the sum of the equilibrium and non-equilibrium moduli pertaining to the molecular relaxation. At the limit of highest frequency, the storage modulus corresponds to the glassy-state modulus, which is about two orders of magnitude larger than the rubbery-state modulus and approaches the sum of the equilibrium modulus and both of the non-equilibrium moduli. We can observe a smooth transition between the rubbery and glassy states as typically observed in dynamic mechanical analysis (DMA) or small amplitude oscillatory shear (SAOS) tests for these materials. In the intermediate frequency range, loss modulus, $E^{\prime\prime}$, is larger than the storage modulus, which indicates significant viscous dissipation is occurring, which is corroborated by the magnitude of the loss factor, $\tan \delta$. At the limit of the highest frequency, the loss factor, $\tan \delta$, approaches zero, indicating elastic response is dominant in that regime. A shortcoming of our proposed model is that it relies on the resistance to molecular relaxation, $G_{\neql(1)}$, to capture the residual strain upon unloading. Since polyborosiloxane (PBS) showed large residual strain in the experiments performed by \cite{konaleLargeDeformationModel2023}, our calibrated non-equilibrium modulus for molecular relaxation, $G_{\neql(1)}$, is a few orders of magnitude larger than the equilibrium shear modulus, $G_{\eq}$. This large value of $G_{\neql(1)}$ is responsible for the larger rubbery-state modulus at low oscillation frequency we see in Figure \ref{fig:oscillation-example} as well as the large loss factor, $\tan \delta$, we obtained in our computation.

\subsection{Significance of the proposed constitutive model}

The constitutive model most closely aligned with this work, addressing strain-rate-induced glass transition, was proposed by \cite{konaleLargeDeformationModel2023}, whose low and high strain-rate experimental data for PBS were used in this study. Their constitutive model is partly integral, representing the chain attachment-detachment kinetics of PBS, and partly differential, representing the viscoplastic flow at high strain rate in a phenomenological way. Although their model was able to capture rate stiffening, it failed to predict the yield and subsequent hardening. In contrast, our proposed model employs a unified differential evolution law-based framework, which is straightforward to implement and capable of capturing the key features of glass transition in elastomers deformed at high strain rates.

\section{Conclusions}
\label{sec:conclusions}

The present work introduces a mechanism‑based finite viscoelastic constitutive model that describes the finite deformation behavior exhibited by amorphous elastomers from the quasi‑static to high strain rate regime. Two underlying mechanisms, molecular relaxation of the free polymer chains in the rubbery state and intermolecular rearrangement of the frozen chains in the glassy state, have been considered in a continuum thermodynamic framework with viscous strain as the internal variable. The differential form of the internal variable evolution laws offers a secondary advantage of developing an efficient computational algorithm that can be easily incorporated into standard finite element solvers. Our proposed constitutive model contains a set of material parameters that bear clear physical meaning, which can be easily identified from standard experimental data sets with minimal effort as demonstrated in this work. Through numerical examples, we were able to demonstrate the essential features of our proposed model, such as being able to capture rate-dependency in both loading and unloading, stabilization of dissipated energy in cyclic loading, and a smooth glass transition behavior when a frequency sweep was applied. However, the model is not without limitations: it does not include separate state variables to account for residual strain, which is commonly observed in many viscoelastic elastomeric materials. Additionally, it neglects the effect of temperature and assumes a simple isotropic bulk viscosity. In the future, the framework may be extended to include thermo‑mechanical coupling, pressure‑dependent volumetric response, and damage evolution without altering its core structure.

In summary, the model bridges a long‑standing gap in polymer mechanics literature by providing a thermodynamically-consistent, physically-based constitutive description from quasi‑static to high strain rate dynamic deformation regimes, taking high strain rate-induced glass transition into account. Its ability to capture the glass transition with a modest number of physically meaningful parameters makes it particularly well-suited for the design‑by‑analysis of architected metamaterials and impact‑absorbing elastomeric components fabricated by additive manufacturing, where accurate strain‑rate sensitivity is crucial. Future work will focus on incorporating thermal effects, pressure‑dependent bulk viscosity, and continuum damage to further broaden the applicability of the constitutive model to analyze high‑performance soft elastomeric structures.

\section*{Data statement}

The code and data required to reproduce the findings reported in this work can be made available upon request to the authors.

\section*{CRediT author statement}

\textbf{Bibekananda Datta:} Conceptualization, Methodology, Software, Validation, Formal analysis, Data Curation, Visualization, Writing - Original Draft.
\textbf{Sushan Nakarmi:} Conceptualization, Investigation, Funding acquisition, Writing - Review \& Editing.
\textbf{Nitin P. Daphalapurkar:} Conceptualization, Investigation, Project administration, Resources, Funding acquisition, Supervision, Writing - Review \& Editing.

\section*{Declaration of competing interest}

The authors declare that they have no known competing financial interests or personal relationships
that could have appeared to influence the work reported in this article.

\section*{Acknowledgment}

This work was performed with support from the following programs: Dynamic Materials Properties (C2) Science Campaign and Advanced Simulation and Computing (ASC)--PEM at Los Alamos National Laboratory, operated by Triad National Security, LLC, for the National Nuclear Security Administration of the U.S. Department of Energy (Contract No. 89233218CNA000001). B.D. would also like to acknowledge the technical discussions on finite viscoelasticity with Dr. Beijun Shen (Columbia University) and Michael Lapera (Johns Hopkins University).

%%%%% BIBLIOGRAPHY
\bibliographystyle{elsarticle-harv} 
\bibliography{finite_viscoelastic_revised}

\begin{appendix}

\section{Numerical implementation procedure}
\label{sec:numerical-implementation-procedure}

Following \cite{reeseTheoryFiniteViscoelasticity1998}, a predictor-corrector type explicit time integration procedure to perform constitutive calculation of our proposed model is given below:

\begin{enumerate}[leftmargin=0.5 cm, itemsep=0pt, topsep=0pt]
    \item \textbf{Compute the kinematic state:} Compute the following kinematic tensors at the material point at the current time, $t + \Delta t$:
    \begin{equation}
    \begin{aligned}
        & J_{t+\Delta t}              &&= \det ( \vect{F}_{t+\Delta t} ), \\
        & \vect{C}_{t+\Delta t}       &&= (\vect{F}^{\T})_{t+\Delta t} (\vect{F})_{t+\Delta t}, \\
        & \vect{b}_{t+\Delta t}       &&= (\vect{F})_{t+\Delta t} (\vect{F}^{\T})_{t+\Delta t}, \\
        & \barVect{F}_{t+\Delta t}    &&= J^{-1/3}_{t+\Delta t} \ \vect{F}_{t+\Delta t}, \\
        & \barVect{C}_{t+\Delta t}    &&= J^{-2/3}_{t+\Delta t} \ \vect{C}_{t+\Delta t}, \\
        & \barVect{b}_{t+\Delta t}    &&= J^{-2/3}_{t+\Delta t} \ \vect{b}_{t+\Delta t}.
    \end{aligned}
    \end{equation}

    \item \textbf{Equilibrium stress:} Use the total deformation tensors from Step 1 to compute the equilibrium stress, $\vsigma_{\eq, t + \Delta t} = \hat{\vsigma}_{\eq} (\barVect{b}_{t+\Delta t}, J_{t + \Delta t})$ using Eq. \eqref{eq:equilibrium-cauchy-stress}.

    \item \textbf{Predictor step:} Assume viscous flow is frozen, \emph{i.e.}, $ \dot{\vect{C}}^v_k = 0 \Rightarrow  {\vect{C}^v_{k, t}} =  {\vect{C}^v_{k, t+\Delta t}} $, and compute a trial state,  $\vect{b}^e_{k, \text{trial}}$, as follows:
    \begin{equation}
    \begin{aligned}
        &\vect{b}^e_{k, \text{trial}}     && = \vect{F}_{t + \Delta t} \left( {\vect{C}^v_{k, t}} \right)^{-1} \vect{F}^{\T}_{t+\Delta t}, \\
        & J^e_{k, \text{trial}}            && = \sqrt{\det \left( \vect{b}^e_{k, \text{trial}}  \right) }, \\
        & \hat{\boldsymbol{\upvarepsilon}}^e_{k, \text{trial}} && = \frac{1}{2} \log \left( \vect{b}^e_{k, \text{trial}} \right),
    \end{aligned}
    \end{equation}
    where, $\hat{\boldsymbol{\upvarepsilon}}^e_{k, \text{trial}} = \ln (\vect{V}^e_{k, \text{trial}}) = \frac{1}{2} \ln (\vect{b}^e_{k, \text{trial}})$ is defined as the trial spatial elastic Hencky strain tensor.

    \item \textbf{Compute trial non-equilibrium stress:} Compute the trial non-equilibrium Cauchy stress using Eq. \eqref{eq:non-equilibrium-cauchy-stress} as follows: \\
    \begin{equation}
        \vsigma_{\neql (k), \text{trial}} = \hat{\vsigma}_{\neql (k)} \left( \barVect{b}^e_{k, \text{trial}},  J^e_{k, \text{trial}} \right).
    \end{equation}

    \item \textbf{Corrector step:} We assume the spatial velocity gradient is zero, $\vect{l} = \vect{0}$. Thus, the Lie time derivative of $\vect{b}^e_k$ reduces as follows:
    \begin{equation}
        \mathcal{L}_v \vect{b}^e_k = \dot{\vect{b}}^e_k 
        - \underbrace{\vect{l} \vect{b}^e_k - \vect{b}^e_k \vect{l}^{\T}}_{\vect{0}} 
        = \dot{\vect{b}}^e_k.
    \end{equation}
    Thus, by using the trial kinematic state and trial non-equilibrium Kirchhoff stress, the viscous stretch evolution law Eq. \eqref{eq:stretch-rate-evolution-law} can be expressed as:
    \begin{equation}
        \dot{\vect{b}}^e_k
        = - 2 \left( \dot{\gamma}^{v,\dev}_k \frac{\vtau_{\neql(k), \text{trial}}^{\dev}}{||\vtau_{\neql(k), \text{trial}}^{\dev}||} + \frac{\dot{\epsilon}^{v,\vol}_k}{3}  \mathds{1}  \right) \vect{b}^e_{k, \text{trial}}.
    \end{equation}
   Using the exponential map technique, the above viscous stretch rate evolution law can be numerically integrated as: 
    \begin{equation} 
    \begin{aligned}
        & \vect{b}^e_{k, t + \Delta t} 
        && = \exp \left[ - 2 \Delta t  \left( \dot{\gamma}^{v,\dev}_k \frac{\vtau_{\neql(k),{\text{trial}}}^{\dev}}{|| \overline{\vtau}_{\neql(k),\text{trial}}^{\dev}||} 
        + \frac{\dot{\epsilon}^{v,\vol}_k}{3}  \mathds{1}  \right) \right]
        \vect{b}^e_{k, \text{trial}}, \\
        \Rightarrow 
        & \left( {\vect{V}^e_{k, t + \Delta t}} \right)^2
        && = \exp \left[ - 2 \Delta t  \left( \dot{\gamma}^{v,\dev}_k \frac{\vtau_{\neql(k),{\text{trial}}}^{\dev}}{|| \overline{\vtau}_{\neql(k),\text{trial}}^{\dev}||} 
        + \frac{\dot{\epsilon}^{v,\vol}_k}{3}  \mathds{1}  \right) \right]
        \left( \vect{V}^e_{k,\text{trial}} \right)^2, \\
        \Rightarrow 
        & \ln  \left( {\vect{V}^e_{k, t + \Delta t}} \right) 
        && =  -\Delta t  \left( \dot{\gamma}^{v,\dev}_k \frac{\vtau_{\neql(k),{\text{trial}}}^{\dev}}{|| \overline{\vtau}_{\neql(k),\text{trial}}^{\dev}||} 
        + \frac{\dot{\epsilon}^{v,\vol}_k}{3}  \mathds{1}  \right)
        + \ln \left( \vect{V}^e_{k, \text{trial}} \right), \\
        \Rightarrow
        &  \hat{\boldsymbol{\upvarepsilon}}^e_{k, t + \Delta t}
        && = \hat{\boldsymbol{\upvarepsilon}}^e_{k, \text{trial}} 
        -  \left( \dot{\gamma}^{v,\dev}_k \frac{\vtau_{\neql(k),{\text{trial}}}^{\dev}}{|| \overline{\vtau}_{\neql(k),\text{trial}}^{\dev}||} 
        + \frac{\dot{\epsilon}^{v,\vol}_k}{3}  \mathds{1}  \right) \Delta t.
    \end{aligned}
    \end{equation}
    Here, the effective deviatoric viscous stretch rate, $\dot{\gamma}^{v,\dev}_k$, is computed either from Eq. \eqref{eq:bergstrom-boyce-law-redefined} or \eqref{eq:ree-eyring-law} and volumetric viscous stretch rate, $\dot{\epsilon}^{v,\vol}_k$, is computed using Eq. \eqref{eq:volumetric-creep-law}.

    \item \textbf{Update state variable:} Compute the following kinematic quantities and update the state variable, $\vect{C}^v_{k, t + \Delta t} $, as follows:
    \begin{equation}
    \begin{aligned}
        & \vect{V}^e_{k, t + \Delta t} 
        &&= \exp \left( \hat{\boldsymbol{\upvarepsilon}}^e_{k, t + \Delta t} \right), \\
        &  \vect{b}^e_{k, t + \Delta t} 
        && = \left( \vect{V}^e_{k, t + \Delta t} \right)^2, \\
        &\vect{C}^v_{k, t + \Delta t} 
        && = \vect{F}^{\T}_{t + \Delta t} \left( {\vect{b}^e_{k, t + \Delta t}} \right)^{-1} \vect{F}_{t + \Delta t}.
    \end{aligned}
    \end{equation}
    In evaluating $\vect{C}^v_k$, we utilized the kinematic relation $ \vect{b}^e_k = \vect{F} {\vect{C}^v_k}^{-1} \vect{F}^{\T}$.

    \item \textbf{Compute final non-equilibrium Cauchy stress:}  Compute the final non-equilibrium Cauchy stress using Eq. \eqref{eq:non-equilibrium-cauchy-stress} as follows:
    \begin{equation}
        \vsigma_{\neql (k), t+\Delta t} = \hat{\vsigma}_{\neql (k)} \left( \barVect{b}^e_{k, t+\Delta t},  J^e_{k, t+\Delta t} \right),
    \end{equation}
    where the elastic volume change is computed as: $J^e_{k, t + \Delta t} = \sqrt{ \det \left( \vect{b}^e_{k, t + \Delta t} \right) }$.

    \item \textbf{Compute total Cauchy stress:} Compute the total Cauchy stress at the current time step: 
    \begin{equation}
        \vsigma_{t+\Delta t} = \vsigma_{\eq, t+\Delta t} + \sum_{k=1}^n \vsigma_{\neql (k), t+\Delta t}.
    \end{equation}
\end{enumerate}
Steps 3--7 are repeated for each non-equilibrium mechanism. The logarithm and exponential of a symmetric second-order tensor, \emph{i.e.}, a matrix, can be computed by eigen decomposition followed by diagonalization.

\end{appendix}

\end{document}